\documentclass[pdflatex,sn-vancouver-num,referee]{sn-jnl}

\usepackage{graphicx}
\usepackage{multirow}
\usepackage{amsmath,amssymb,amsfonts}
\usepackage{amsthm}
\usepackage[title]{appendix}
\usepackage{xcolor}
\usepackage{textcomp}
\usepackage{manyfoot}
\usepackage{booktabs}
\usepackage{algorithm}
\usepackage{algorithmicx}
\usepackage{algpseudocode}
\usepackage{listings}
\usepackage{fvextra}
\usepackage{mdframed}
\usepackage{pgf-pie}
\usepackage{pgfplots}
\usepackage{longtable}
\usepackage{array}
\usepackage{comment}
\usepackage{hyperref}
\usepackage{pdfpages}

\pgfplotsset{compat=1.18}
\usepackage{tikz}
\DeclareUnicodeCharacter{2265}{$\geq$}
\DeclareUnicodeCharacter{2264}{$\leq$}
\DeclareUnicodeCharacter{22C6}{$\star$}

\usepackage[T1]{fontenc}
\usepackage{lmodern}
\hypersetup{bookmarksdepth=section}
\usepackage{enumitem}
\usepackage{needspace}
\usepackage{xcolor}
\usepackage[most]{tcolorbox}

\definecolor{qblue}{RGB}{34,76,152}
\definecolor{qlight}{RGB}{245,247,250}
\definecolor{qborder}{RGB}{210,218,230}

\definecolor{aboxbg}{RGB}{252,252,253}
\definecolor{aboxborder}{RGB}{225,229,236}

\definecolor{sectionbg}{RGB}{232,238,248}
\definecolor{sectionborder}{RGB}{180,194,216}

\definecolor{answerlabelblue}{RGB}{74,110,170}

\newtcolorbox{sectionbox}{
  enhanced,
  breakable,
  colback=sectionbg,
  colframe=sectionborder,
  boxrule=0.8pt,
  arc=2mm,
  left=2mm,
  right=2mm,
  top=1.4mm,
  bottom=1.4mm,
  fontupper=\small,
  before upper=\raggedright
}

\newtcolorbox{questionbox}{
  enhanced,
  breakable,
  colback=qlight,
  colframe=qborder,
  boxrule=0.6pt,
  arc=2mm,
  left=2mm,
  right=2mm,
  top=1.2mm,
  bottom=1.2mm,
  fontupper=\small,
  before upper=\raggedright
}

\newtcolorbox{answerbox}{
  enhanced,
  breakable,
  colback=aboxbg,
  colframe=aboxborder,
  boxrule=0.5pt,
  arc=1.5mm,
  left=2mm,
  right=2mm,
  top=0.8mm,
  bottom=0.5mm,
  fontupper=\small,
  before upper=\raggedright
}

\newcommand{\appqsection}[1]{%
  \vspace{0.2em}
  \begin{sectionbox}
  \textbf{\textcolor{qblue}{#1}}
  \end{sectionbox}
}

\newcommand{\question}[1]{%
  \begin{questionbox}
  \textbf{\textcolor{qblue}{Question:}} #1
  \end{questionbox}
}

\newcommand{\answertype}[1]{%
  \begin{answerbox}
  \textit{\textcolor{answerlabelblue}{Answer:}}%
  \makebox[\dimexpr\linewidth-4.5em\relax][c]{#1}
  \end{answerbox}

}

\newcommand{\possanswers}[1]{%
  \begin{answerbox}
  \textit{\textcolor{answerlabelblue}{Answers:}} #1
  \end{answerbox}
}

\newcommand{\inlineanswers}[1]{%
  \begin{answerbox}
  \noindent\textit{\textcolor{answerlabelblue}{Answers:}}%
  \hfill #1 \hfill\mbox{}
  \end{answerbox}
}

\newcommand{\fivepointscale}[2]{%
  \begin{answerbox}
  \noindent\textit{\textcolor{answerlabelblue}{Answers:}}%
  \hfill
  {\footnotesize #1}\hspace{2em}%
  \textbf{1}\hspace{1.3em}%
  \textbf{2}\hspace{1.3em}%
  \textbf{3}\hspace{1.3em}%
  \textbf{4}\hspace{1.3em}%
  \textbf{5}\hspace{2em}%
  {\footnotesize #2}%
  \hfill\mbox{}
  \end{answerbox}
}

\setlist[itemize]{
  leftmargin=1.6em,
  itemsep=0.2em,
  topsep=0.2em,
  parsep=0pt,
  partopsep=0pt,
  label=--
}

\usepackage{fvextra}
\usepackage[most]{tcolorbox}

\newtcolorbox{promptbox}{
  enhanced,
  breakable,
  colback=aboxbg,
  colframe=aboxborder,
  boxrule=0.5pt,
  arc=1.5mm,
  left=1mm,
  right=1mm,
  top=1mm,
  bottom=1mm,
  fonttitle=\bfseries,
  coltitle=qblue,
}
\theoremstyle{thmstyleone}

\theoremstyle{thmstyletwo}

\theoremstyle{thmstylethree}

\expandafter\def\csname statLCn\endcsname{30}
\expandafter\def\csname statLCuseful\endcsname{22}
\expandafter\def\csname pctLCuseful\endcsname{73\%}
\expandafter\def\csname statLClessuseful\endcsname{13}
\expandafter\def\csname pctLClessuseful\endcsname{43\%}
\expandafter\def\csname statLCtrustless\endcsname{16}
\expandafter\def\csname pctLCtrustless\endcsname{53\%}
\expandafter\def\csname statLCaionly\endcsname{25}
\expandafter\def\csname pctLCaionly\endcsname{83\%}
\expandafter\def\csname statLChumanonly\endcsname{25}
\expandafter\def\csname pctLChumanonly\endcsname{83\%}
\expandafter\def\csname statLCalignedpoor\endcsname{15}
\expandafter\def\csname pctLCalignedpoor\endcsname{50\%}
\expandafter\def\csname statLCanyerror\endcsname{20}
\expandafter\def\csname pctLCanyerror\endcsname{67\%}
\expandafter\def\csname statLCminor\endcsname{15}
\expandafter\def\csname pctLCminor\endcsname{50\%}
\expandafter\def\csname statLCsevere\endcsname{5}
\expandafter\def\csname pctLCsevere\endcsname{17\%}
\expandafter\def\csname statLCincorporated\endcsname{25}
\expandafter\def\csname pctLCincorporated\endcsname{83\%}
\expandafter\def\csname statLCfull\endcsname{2}
\expandafter\def\csname pctLCfull\endcsname{7\%}
\expandafter\def\csname statLCsupplement\endcsname{24}
\expandafter\def\csname pctLCsupplement\endcsname{80\%}
\expandafter\def\csname statLCnotice\endcsname{26}
\expandafter\def\csname pctLCnotice\endcsname{87\%}
\expandafter\def\csname statLCinternal\endcsname{29}
\expandafter\def\csname pctLCinternal\endcsname{97\%}
\expandafter\def\csname statLCconsent\endcsname{19}
\expandafter\def\csname pctLCconsent\endcsname{63\%}
\expandafter\def\csname statLCprior\endcsname{22}
\expandafter\def\csname pctLCprior\endcsname{73\%}
\expandafter\def\csname statLChumanclear\endcsname{17}
\expandafter\def\csname pctLChumanclear\endcsname{57\%}
\expandafter\def\csname statLCequalclear\endcsname{10}
\expandafter\def\csname pctLCequalclear\endcsname{33\%}
\expandafter\def\csname statLCequaltrust\endcsname{10}
\expandafter\def\csname pctLCequaltrust\endcsname{33\%}
\expandafter\def\csname statLCequaluseful\endcsname{10}
\expandafter\def\csname pctLCequaluseful\endcsname{33\%}
\expandafter\def\csname statLChumanmixed\endcsname{4}
\expandafter\def\csname pctLChumanmixed\endcsname{13\%}
\expandafter\def\csname statLChumanuseful\endcsname{25}
\expandafter\def\csname pctLChumanuseful\endcsname{83\%}
\expandafter\def\csname statLCsupervision\endcsname{24}
\expandafter\def\csname pctLCsupervision\endcsname{80\%}
\expandafter\def\csname statLCmajorconcerns\endcsname{3}
\expandafter\def\csname pctLCmajorconcerns\endcsname{10\%}
\expandafter\def\csname statLCcomfortable\endcsname{24}
\expandafter\def\csname pctLCcomfortable\endcsname{80\%}
\expandafter\def\csname statCVn\endcsname{27}
\expandafter\def\csname statCVuseful\endcsname{26}
\expandafter\def\csname pctCVuseful\endcsname{96\%}
\expandafter\def\csname statCVlessuseful\endcsname{12}
\expandafter\def\csname pctCVlessuseful\endcsname{44\%}
\expandafter\def\csname statCVtrustless\endcsname{17}
\expandafter\def\csname pctCVtrustless\endcsname{63\%}
\expandafter\def\csname statCVaionly\endcsname{20}
\expandafter\def\csname pctCVaionly\endcsname{74\%}
\expandafter\def\csname statCVhumanonly\endcsname{21}
\expandafter\def\csname pctCVhumanonly\endcsname{78\%}
\expandafter\def\csname statCValignedpoor\endcsname{9}
\expandafter\def\csname pctCValignedpoor\endcsname{33\%}
\expandafter\def\csname statCVanyerror\endcsname{19}
\expandafter\def\csname pctCVanyerror\endcsname{70\%}
\expandafter\def\csname statCVminor\endcsname{15}
\expandafter\def\csname pctCVminor\endcsname{56\%}
\expandafter\def\csname statCVsevere\endcsname{4}
\expandafter\def\csname pctCVsevere\endcsname{15\%}
\expandafter\def\csname statCVincorporated\endcsname{22}
\expandafter\def\csname pctCVincorporated\endcsname{81\%}
\expandafter\def\csname statCVfull\endcsname{1}
\expandafter\def\csname pctCVfull\endcsname{4\%}
\expandafter\def\csname statCVsupplement\endcsname{27}
\expandafter\def\csname pctCVsupplement\endcsname{100\%}
\expandafter\def\csname statCVnotice\endcsname{25}
\expandafter\def\csname pctCVnotice\endcsname{93\%}
\expandafter\def\csname statCVinternal\endcsname{26}
\expandafter\def\csname pctCVinternal\endcsname{96\%}
\expandafter\def\csname statCVconsent\endcsname{25}
\expandafter\def\csname pctCVconsent\endcsname{93\%}
\expandafter\def\csname statCVprior\endcsname{12}
\expandafter\def\csname pctCVprior\endcsname{44\%}
\expandafter\def\csname statCVhumanclear\endcsname{11}
\expandafter\def\csname pctCVhumanclear\endcsname{41\%}
\expandafter\def\csname statCVequalclear\endcsname{11}
\expandafter\def\csname pctCVequalclear\endcsname{41\%}
\expandafter\def\csname statCVequaltrust\endcsname{8}
\expandafter\def\csname pctCVequaltrust\endcsname{30\%}
\expandafter\def\csname statCVequaluseful\endcsname{11}
\expandafter\def\csname pctCVequaluseful\endcsname{41\%}
\expandafter\def\csname statCVhumanmixed\endcsname{9}
\expandafter\def\csname pctCVhumanmixed\endcsname{33\%}
\expandafter\def\csname statCVhumanuseful\endcsname{18}
\expandafter\def\csname pctCVhumanuseful\endcsname{67\%}
\expandafter\def\csname statCVsupervision\endcsname{24}
\expandafter\def\csname pctCVsupervision\endcsname{89\%}
\expandafter\def\csname statCVmajorconcerns\endcsname{0}
\expandafter\def\csname pctCVmajorconcerns\endcsname{0\%}
\expandafter\def\csname statCVcomfortable\endcsname{17}
\expandafter\def\csname pctCVcomfortable\endcsname{63\%}
\expandafter\def\csname statLNn\endcsname{35}
\expandafter\def\csname statLNuseful\endcsname{34}
\expandafter\def\csname pctLNuseful\endcsname{97\%}
\expandafter\def\csname statLNlessuseful\endcsname{8}
\expandafter\def\csname pctLNlessuseful\endcsname{23\%}
\expandafter\def\csname statLNtrustless\endcsname{11}
\expandafter\def\csname pctLNtrustless\endcsname{31\%}
\expandafter\def\csname statLNaionly\endcsname{31}
\expandafter\def\csname pctLNaionly\endcsname{89\%}
\expandafter\def\csname statLNhumanonly\endcsname{29}
\expandafter\def\csname pctLNhumanonly\endcsname{83\%}
\expandafter\def\csname statLNalignedpoor\endcsname{8}
\expandafter\def\csname pctLNalignedpoor\endcsname{23\%}
\expandafter\def\csname statLNanyerror\endcsname{19}
\expandafter\def\csname pctLNanyerror\endcsname{54\%}
\expandafter\def\csname statLNminor\endcsname{15}
\expandafter\def\csname pctLNminor\endcsname{43\%}
\expandafter\def\csname statLNsevere\endcsname{4}
\expandafter\def\csname pctLNsevere\endcsname{11\%}
\expandafter\def\csname statLNincorporated\endcsname{35}
\expandafter\def\csname pctLNincorporated\endcsname{100\%}
\expandafter\def\csname statLNfull\endcsname{10}
\expandafter\def\csname pctLNfull\endcsname{29\%}
\expandafter\def\csname statLNsupplement\endcsname{34}
\expandafter\def\csname pctLNsupplement\endcsname{97\%}
\expandafter\def\csname statLNnotice\endcsname{32}
\expandafter\def\csname pctLNnotice\endcsname{91\%}
\expandafter\def\csname statLNinternal\endcsname{34}
\expandafter\def\csname pctLNinternal\endcsname{97\%}
\expandafter\def\csname statLNconsent\endcsname{29}
\expandafter\def\csname pctLNconsent\endcsname{83\%}
\expandafter\def\csname statLNprior\endcsname{27}
\expandafter\def\csname pctLNprior\endcsname{77\%}
\expandafter\def\csname statLNhumanclear\endcsname{20}
\expandafter\def\csname pctLNhumanclear\endcsname{57\%}
\expandafter\def\csname statLNequalclear\endcsname{13}
\expandafter\def\csname pctLNequalclear\endcsname{37\%}
\expandafter\def\csname statLNequaltrust\endcsname{20}
\expandafter\def\csname pctLNequaltrust\endcsname{57\%}
\expandafter\def\csname statLNequaluseful\endcsname{16}
\expandafter\def\csname pctLNequaluseful\endcsname{46\%}
\expandafter\def\csname statLNhumanmixed\endcsname{1}
\expandafter\def\csname pctLNhumanmixed\endcsname{3\%}
\expandafter\def\csname statLNhumanuseful\endcsname{34}
\expandafter\def\csname pctLNhumanuseful\endcsname{97\%}
\expandafter\def\csname statLNsupervision\endcsname{28}
\expandafter\def\csname pctLNsupervision\endcsname{80\%}
\expandafter\def\csname statLNmajorconcerns\endcsname{1}
\expandafter\def\csname pctLNmajorconcerns\endcsname{3\%}
\expandafter\def\csname statLNcomfortable\endcsname{25}
\expandafter\def\csname pctLNcomfortable\endcsname{71\%}
\expandafter\def\csname statAlln\endcsname{92}
\expandafter\def\csname statAlluseful\endcsname{82}
\expandafter\def\csname pctAlluseful\endcsname{89\%}
\expandafter\def\csname statAlllessuseful\endcsname{33}
\expandafter\def\csname pctAlllessuseful\endcsname{36\%}
\expandafter\def\csname statAlltrustless\endcsname{44}
\expandafter\def\csname pctAlltrustless\endcsname{48\%}
\expandafter\def\csname statAllaionly\endcsname{76}
\expandafter\def\csname pctAllaionly\endcsname{83\%}
\expandafter\def\csname statAllhumanonly\endcsname{75}
\expandafter\def\csname pctAllhumanonly\endcsname{82\%}
\expandafter\def\csname statAllalignedpoor\endcsname{32}
\expandafter\def\csname pctAllalignedpoor\endcsname{35\%}
\expandafter\def\csname statAllanyerror\endcsname{58}
\expandafter\def\csname pctAllanyerror\endcsname{63\%}
\expandafter\def\csname statAllminor\endcsname{45}
\expandafter\def\csname pctAllminor\endcsname{49\%}
\expandafter\def\csname statAllsevere\endcsname{13}
\expandafter\def\csname pctAllsevere\endcsname{14\%}
\expandafter\def\csname statAllincorporated\endcsname{82}
\expandafter\def\csname pctAllincorporated\endcsname{89\%}
\expandafter\def\csname statAllfull\endcsname{13}
\expandafter\def\csname pctAllfull\endcsname{14\%}
\expandafter\def\csname statAllsupplement\endcsname{85}
\expandafter\def\csname pctAllsupplement\endcsname{92\%}
\expandafter\def\csname statAllnotice\endcsname{83}
\expandafter\def\csname pctAllnotice\endcsname{90\%}
\expandafter\def\csname statAllinternal\endcsname{89}
\expandafter\def\csname pctAllinternal\endcsname{97\%}
\expandafter\def\csname statAllconsent\endcsname{73}
\expandafter\def\csname pctAllconsent\endcsname{79\%}
\expandafter\def\csname statAllprior\endcsname{61}
\expandafter\def\csname pctAllprior\endcsname{66\%}
\expandafter\def\csname statAllhumanclear\endcsname{48}
\expandafter\def\csname pctAllhumanclear\endcsname{52\%}
\expandafter\def\csname statAllequalclear\endcsname{34}
\expandafter\def\csname pctAllequalclear\endcsname{37\%}
\expandafter\def\csname statAllequaltrust\endcsname{38}
\expandafter\def\csname pctAllequaltrust\endcsname{41\%}
\expandafter\def\csname statAllequaluseful\endcsname{37}
\expandafter\def\csname pctAllequaluseful\endcsname{40\%}
\expandafter\def\csname statAllhumanmixed\endcsname{14}
\expandafter\def\csname pctAllhumanmixed\endcsname{15\%}
\expandafter\def\csname statAllhumanuseful\endcsname{77}
\expandafter\def\csname pctAllhumanuseful\endcsname{84\%}
\expandafter\def\csname statAllsupervision\endcsname{76}
\expandafter\def\csname pctAllsupervision\endcsname{83\%}
\expandafter\def\csname statAllmajorconcerns\endcsname{4}
\expandafter\def\csname pctAllmajorconcerns\endcsname{4\%}
\expandafter\def\csname statAllcomfortable\endcsname{66}
\expandafter\def\csname pctAllcomfortable\endcsname{72\%}
\newcommand{\stat}[2]{\csname stat#1#2\endcsname}
\newcommand{\pc}[2]{\csname pct#1#2\endcsname}
\newcommand{\result}[2]{\stat{#1}{#2}/\stat{#1}{n} (\pc{#1}{#2})}

\expandafter\def\csname participationLC\endcsname{31/89 (35\%)}
\expandafter\def\csname participationCV\endcsname{27/70 (39\%)}
\expandafter\def\csname participationLN\endcsname{35/82 (43\%)}
\expandafter\def\csname participationAll\endcsname{93/241 (39\%)}
\newcommand{\participation}[1]{\csname participation#1\endcsname}

\begin{document}

\title[Authors' Response to AI-Based Reviews:\\A Mixed-Methods Pilot Study]
{\centering Authors' Response to AI-Based Reviews:\\A Mixed-Methods Pilot Study}

\author*[1,2]{\fnm{César} \sur{Leblanc}}\email{phd.cesar.leblanc@gmail.com}
\equalcont{These authors contributed equally to this work.}
\author*[3,4]{\fnm{Lukáš} \sur{Picek}}\email{lukaspicek@gmail.com}
\equalcont{These authors contributed equally to this work.}
\affil[1]{\orgname{École Normale Supérieure}, \orgaddress{\city{Paris}, \country{France}}}
\affil[2]{\orgname{Sorbonne University}, \orgaddress{\city{Paris}, \country{France}}}
\affil[3]{\orgname{University of West Bohemia}, \orgaddress{\city{Plzeň}, \country{Czech Republic}}}
\affil[4]{\orgname{Massachusetts Institute of Technology}, \orgaddress{\city{Cambridge}, \country{USA}}}

\keywords{Large Language Models, Scholarly peer review, ChatGPT, Peer review quality, AI-based review, Research ethics, Author perceptions, Generative AI, Data privacy}


\abstract{
\textbf{Background:}
Large language models are increasingly used to assist scholarly peer review, but there is limited evidence about how authors judge their feedback and use it to revise their papers.

\medskip
\textbf{Methods:}
We studied authors' experiences of receiving an auxiliary AI review alongside human reviews in three computer-science cohorts: LifeCLEF 2025, LifeCLEF 2026, and CVWW 2026. Post-review questionnaires yielded 92 responses concerning 71 papers. We summarized closed-ended answers using descriptive statistics and examined free-text responses through thematic analysis.

\medskip
\textbf{Results:}
Most responses rated the AI review useful (\result{All}{useful}), and \result{All}{incorporated} reported using at least some of its suggestions in revision. Authors reported that AI identified issues missed by human reviewers (\result{All}{aionly}). However, \result{All}{trustless} trusted AI feedback less than human feedback, and \result{All}{anyerror} reported incorrect or misleading comments. Authors valued suggestions that improved clarity and reproducibility, but also described overlooked evidence, inappropriate demands, and difficulty finding useful comments within lengthy reviews. They emphasized the need to check AI feedback and retain control over how their manuscripts were processed. Future review by both humans and AI was supported by \result{All}{supplement}; \result{All}{notice} preferred advance notice and \result{All}{consent} supported explicit consent.

\medskip
\textbf{Conclusions:}
Authors often found AI reviews useful for revision without trusting them as much as human reviews. These exploratory findings favor a supplementary role for AI, with human oversight, clear disclosure, and author consent. They do not establish that AI can replace human reviewers or improve manuscript quality.
}

\maketitle

\section{Background}\label{background}

Peer review is the core mechanism of scientific communication. Yet the feedback authors receive can vary substantially in clarity, depth, constructiveness, and usefulness~\cite{horbach2018changing,aczel2025present,superchi2019tools,chong2024feedback}. Reviews are expected to be fair, specific, thorough, and actionable, but standards for high-quality review reports are difficult to define~\cite{sizo2025defining}. In practice, review quality depends heavily on the individual reviewers assigned to a manuscript, including their expertise, motivation, language proficiency, and interpretation of the review task~\cite{kelly2014peer}. Reviewing also relies heavily on unpaid service labor~\cite{leblanc2023scientific}. Consistent with this variability, agreement between reviewers is often modest in both journal peer review and large machine-learning venues~\cite{rothwell2000reproducibility,neurips2021consistency,kumar2023finding}.

Supplementary feedback could identify ambiguous writing, missing methodological details, or weakly supported claims. Its value depends on usefulness, reliability, accountability, confidentiality, and transparency~\cite{aczel2025present,tennant2020limitations,keserlioglu2019impact}.

Large language models (LLMs), already used across research workflows, are increasingly discussed as tools for supporting peer review. Major venues and publishers have responded with policies or pilot programs for AI-based reviewing~\cite{liang2024monitoring,bhavsar2025policies}, including NeurIPS~\cite{neurips2025llm_policy}, ICML~\cite{icml2026llm_policy}, AAAI~\cite{aaai2025_ai_assisted_review}, and Nature Portfolio~\cite{nature_ai_editorial_policy}. LLMs can generate review-like feedback quickly and at low cost.

However, fluent feedback is not the same as reliable judgment. LLMs can produce plausible but unsupported, incorrect, or internally inconsistent claims, often described as hallucinations~\cite{huang2025survey,anh2025survey}. Reviews must evaluate manuscripts accurately, apply appropriate standards, and avoid inventing problems or overstating weaknesses. AI-based feedback can therefore be useful in some respects while still creating extra work for authors if it is poorly grounded or calibrated.

Data confidentiality is a further concern. When an unpublished manuscript is submitted to a peer-review workflow that includes a third-party AI system, authors face risks distinct from those of inaccurate feedback: their confidential and unpublished work may be exposed to additional processing, retention, or disclosure risks beyond the review itself \cite{cheng2024generative,helmy2025ten}. These concerns are especially salient because AI use in peer review remains governed by inconsistent venue policies, with many journals prohibiting it and others allowing it only under confidentiality conditions \cite{leung2023best,li2024use}.

Recent work has framed AI in peer review as both a practical opportunity and a governance challenge~\cite{checco2021ai,hosseini2023fighting,lee2025role,kocak2025ensuring,hoyt2025generative,perlis2025artificial}. Empirical studies have begun to examine specific uses of AI in scholarly publishing, including reviewer identification and diversity in peer-review recruitment~\cite{teixeira2025ai}, publisher policies on AI chatbot use~\cite{bhavsar2025policies}, and editors' attitudes toward AI chatbots in editorial workflows~\cite{ng2025attitudes}. Much of this discussion focuses on reviewers, editors, publishers, or policy.

Authors are still underrepresented in this debate. They are not just recipients of editorial decisions; they are the people who read reviews, judge which comments are credible, and decide what to use in revision. Early evidence suggests that authors may be more comfortable using AI as a pre-submission self-check than as a replacement for expert review~\cite{lemberger2026authors}. Evidence remains limited about how these experiences and governance expectations vary across live review settings. Do they find the feedback useful? Do they trust it? Does it raise issues human reviewers missed? Do authors use it when revising their papers? And what notice or consent do they expect when unpublished manuscripts are processed by third-party AI systems? We address these questions by examining authors' responses after they received an auxiliary AI-based review alongside standard human peer reviews in three computer science pilot cohorts: LifeCLEF 2025, LifeCLEF 2026, and CVWW 2026.

\section{Methods}\label{methods}

\noindent\textbf{Study design and setting.}
We conducted an exploratory mixed-methods study of authors' experiences of auxiliary AI-based reviews in three computer-science cohorts: LifeCLEF 2025~\cite{lifeclef2025web,picek2025overviewa}, LifeCLEF 2026~\cite{lifeclef2026web}, and CVWW 2026~\cite{cvww2026web}. The two LifeCLEF cohorts concern successive editions of the same biodiversity benchmark initiative within CLEF. Their working-notes papers describe systems submitted to competition tasks. CVWW is a regional computer-vision workshop emphasizing developmental feedback. The design combines descriptive questionnaire results with thematic interpretation of free-text accounts; it does not compare prompts or models experimentally.

\bigskip
\noindent\textbf{Participants and recruitment.}
Authors who had received AI and human reviews were invited by email to complete a voluntary Google Forms questionnaire. Invitations explained the research purpose and treated questionnaire submission as consent to participate. Recruitment focused on authors preparing camera-ready papers. The quantitative sample comprised 30, 35, and 27 responses concerning 20, 31, and 20 papers in LifeCLEF 2025, LifeCLEF 2026, and CVWW 2026, respectively (92 responses; 71 papers). One questionnaire in each of CVWW 2026 and LifeCLEF 2026 explicitly covered two separately invited papers and was counted once per paper. Separate valid questionnaires for the same paper or from the same person were retained; each original free-text answer was considered once. Thus, ``responses'' denotes paper-linked observations, not distinct individuals.

Operational participation was calculated at the author--paper invitation level: \participation{LC}, \participation{LN}, and \participation{CV}, respectively, or \participation{All} pooled. Individuals involved in multiple papers could contribute more than once to both numerator and denominator. Participation included all submissions, including one entirely blank LifeCLEF 2025 form excluded from substantive analysis. Physical-form completion and the sampling frame are detailed in CHERRIES (Appendix~\ref{app:cherries}; Additional file 1). Submissions spanned 8 December 2025--1 February 2026, 13 July--22 August 2026, and 13 January--10 February 2026, respectively. Both LifeCLEF cohorts included larger proportions of undergraduate and master's students than CVWW (Table~\ref{tab:respondents}).

    \begin{table}[htbp]
    \centering
    \caption{\textbf{Self-reported respondent seniority.}}
    \small\setlength{\tabcolsep}{3pt}
    \begin{tabular}{lrrrrrr}
    \toprule
     & Undergrad. & Graduate & PhD & Postdoc & Faculty & Neither\\
    \midrule
    LifeCLEF 2025 & 6 (20\%) & 15 (50\%) & 2 (7\%) & 3 (10\%) & 3 (10\%) & 1 (3\%)\\
LifeCLEF 2026 & 5 (14\%) & 15 (43\%) & 4 (11\%) & 1 (3\%) & 3 (9\%) & 7 (20\%)\\
CVWW 2026 & 0 (0\%) & 3 (11\%) & 9 (33\%) & 4 (15\%) & 9 (33\%) & 2 (7\%)\\
\bottomrule\end{tabular}\par\smallskip\footnotesize Counts (whole percentages): LifeCLEF 2025 $n=30$, LifeCLEF 2026 $n=35$, CVWW 2026 $n=27$. Percentages may not sum to 100\% because of rounding.\label{tab:respondents}\end{table}

\bigskip
\noindent\textbf{AI-review procedure.}
All participating papers received at least two human reviews and one AI review generated in ChatGPT temporary mode. AI feedback was explicitly labeled as auxiliary and did not contribute to acceptance or rejection decisions. Reviews were uploaded to EasyChair for the two LifeCLEF cohorts and to CMT for CVWW. The models were ChatGPT-4o (\texttt{gpt-4o-2024-11-20}), ChatGPT 5.5 (\texttt{gpt-5.5-2026-04-23}), and ChatGPT-5.2 (\texttt{gpt-5.2-2025-12-11}) for LifeCLEF 2025, LifeCLEF 2026, and CVWW 2026, respectively.

LifeCLEF 2025 authors received advance email notification without a separate opt-in or opt-out. CVWW 2026 and LifeCLEF 2026 required affirmative opt-in before manuscript processing; papers without permission were not processed. Permission for AI processing and consent to participate in the survey were separate decisions.

The LifeCLEF 2025 prompt emphasized reproducibility, experimental evaluation, and constructive feedback. Initial outputs were considered overly positive, prompting the follow-up instruction ``Act more like a Reviewer 2.'' The LifeCLEF 2026 prompt specified the working-notes genre and task requirements, competition protocols, reproducibility, evidence anchoring, and reference verification. It supplied task-specific citations and calibrated recommendations to useful system reports rather than archival-paper novelty. The CVWW prompt adopted a demanding top-conference reviewer role, requesting evidence anchors, methodological and novelty assessment, severity labels, and concrete revisions. All prompts are reproduced in Appendix~\ref{app:prompts}.

\bigskip
\noindent\textbf{Questionnaire and outcomes.}
Authors completed substantively equivalent questionnaires assessing seniority, prior AI use, perceived usefulness and trust, comparison with human reviews, complementary issues identified, incorporation of feedback, perceived inaccuracies, future use, and preferences regarding notice, privacy, and consent. The questionnaire was developed for this pilot without an external pilot study or formal psychometric validation. The full instrument is reproduced in Appendix~\ref{app:questionnaire}.

Two optional free-text questions requested examples of incorrect or misleading comments and final reflections. A separate concerns item was categorical. Attribution was optional; names and paper identifiers were excluded from the qualitative analysis table, and published quotations identify only the cohort.

\bigskip
\noindent\textbf{Reflexivity.}\phantomsection\label{reflexivity}
The investigators work in AI and biodiversity research and were involved in organizing the review workflows, designing prompts, and administering questionnaires. Their professional interest in AI assistance may favor positive interpretations, while organizer authority may influence participants' responses. Neither investigator has primary training in qualitative methods. The analysis therefore attends to favorable, critical, and ambivalent accounts, including responses that challenge the questionnaire's comparisons or the proposed role of AI.

\bigskip
\noindent\textbf{Analysis.}
All quantitative summaries were calculated directly from the three raw exports using Python. Counts and whole percentages are reported, with rounding only at display. Ratings 1--2 on the five-point scales indicate lower comparative usefulness, poor alignment, and lower trust. Complementarity uses affirmative answers to the two issues-missed questions. Any inaccuracy comprises all options except ``No, all comments were accurate''; minor inaccuracies use the corresponding option, and clearly incorrect/misleading or irrelevant comments combine the remaining two options. Both severity summaries use the full cohort denominator. Incorporation combines partial and full use; supplementary review means ``Yes, by both humans and AI''; prior self-review combines self-review-only and both writing/self-review; consent combines both affirmative options. Responses receive equal weight in pooling. No inferential tests, nonresponse weights, or power calculation were used. A form-level check without either expansion did not alter the interpretation.

We examined the complete three-cohort free-text corpus through thematic synthesis informed by the six phases of reflexive thematic analysis~\cite{braun2006using}: familiarization, semantic coding, candidate-theme development, review, definition, and reporting. The analysis was primarily inductive within the questionnaire's topic boundaries; themes are interpretive accounts, not objectively measured categories. Sixty-one substantive extracts from 44 questionnaires informed the synthesis. Candidate themes were checked against every extract and compared across cohorts, with attention to divergent accounts and distinctions between author-reported error, disagreement, and relevance. Quotations were selected for analytical value and contrasting views and checked against the raw text. An extract-level coding record documents these interpretations. No inter-rater reliability, member checking, or saturation claim is made. SRQR reporting is provided in Appendix~\ref{app:srqr} (Additional file 2).

Quantitative and qualitative findings were integrated by examining how accounts of usefulness, error, and governance explained or qualified the closed-ended patterns. Cohort contrasts remain descriptive because context, seniority, prompt, model, timing, and procedure co-vary.

\section{Quantitative results}

\noindent\textbf{Usefulness, comparative value, and complementarity.}
Most responses rated the auxiliary AI review useful: \result{All}{useful} overall, with cohort proportions of \pc{LC}{useful}, \pc{LN}{useful}, and \pc{CV}{useful} for LifeCLEF 2025, LifeCLEF 2026, and CVWW 2026, respectively (Fig.~\ref{fig:usefulness_value_complementarity}). Usefulness did not imply equivalent standing with human feedback. Across cohorts, \result{All}{lessuseful} rated AI less useful and \result{All}{trustless} trusted it less than human reviews; midpoint ratings accounted for \result{All}{equaluseful} and \result{All}{equaltrust}, respectively. Human reviews provided clearer suggestions in \result{All}{humanclear}, while \result{All}{equalclear} rated the two sources approximately equal in clarity.

The two sources were widely perceived as complementary: \result{All}{aionly} reported that AI identified issues humans did not, and \result{All}{humanonly} reported the converse. Both patterns were common within each cohort. These judgments concern non-overlapping comments, not independently verified accuracy or relevance.

    \begin{figure}[htbp]
    \centering
    \begin{tikzpicture}
    \begin{axis}[
     xbar, width=0.73\linewidth, height=7.0cm,
     xmin=0, xmax=110, xlabel={Responses (\%)},
     label style={font=\footnotesize}, tick label style={font=\footnotesize},
     xtick={0,20,40,60,80,100}, ytick={1,2,3,4,5},
     yticklabels={{\shortstack[r]{Was the AI-based\\review useful?}},{\shortstack[r]{Was AI-based review less\\useful than human reviews?}},{\shortstack[r]{Was AI-based review trusted\\less than human reviews?}},{\shortstack[r]{Did AI-based review identify\\issues humans did not?}},{\shortstack[r]{Did human reviews identify\\issues AI did not?}}}, y dir=reverse,
     y tick label style={anchor=east,font=\footnotesize,xshift=-2pt},
     bar width=6pt, enlarge y limits=0.17, axis lines*=left,
     legend style={at={(0.5,-0.20)},anchor=north,draw=none,font=\footnotesize},
     legend columns=3,
     xmajorgrids=true, grid style={dashed,gray!40}, point meta=x,
     nodes near coords={\pgfmathprintnumber[fixed,precision=0]{\pgfplotspointmeta}},
     nodes near coords style={font=\scriptsize,xshift=2pt}]
    \addplot[fill=blue!70!black,draw=blue!70!black,bar shift=8.5pt] coordinates {
(73.33333333,1)
(43.33333333,2)
(53.33333333,3)
(83.33333333,4)
(83.33333333,5)
};
\addplot[fill=orange!85!black,draw=orange!85!black,bar shift=0pt] coordinates {
(97.14285714,1)
(22.85714286,2)
(31.42857143,3)
(88.57142857,4)
(82.85714286,5)
};
\addplot[fill=black!75,draw=black!75,bar shift=-8.5pt] coordinates {
(96.29629630,1)
(44.44444444,2)
(62.96296296,3)
(74.07407407,4)
(77.77777778,5)
};
\legend{LifeCLEF 2025 ($n=30$),LifeCLEF 2026 ($n=35$),CVWW 2026 ($n=27$)}
\end{axis}\end{tikzpicture}
\caption{\textbf{Usefulness, comparative value, and complementarity of the AI-based review.} ``Less useful'' and ``trusted less'' combine ratings 1--2. Novel comments indicate perceived complementarity, not verified correctness or usefulness. Sample sizes: LifeCLEF 2025 $n=30$, LifeCLEF 2026 $n=35$, CVWW 2026 $n=27$.}
\label{fig:usefulness_value_complementarity}
\end{figure}

\noindent\textbf{Reported limitations and inaccuracies.}
Across cohorts, \result{All}{anyerror} reported at least some incorrect or misleading comments (Fig.~\ref{fig:limitations_misfit}). Minor inaccuracies accounted for \result{All}{minor}; \result{All}{severe} selected clearly incorrect/misleading or largely irrelevant comments. Poor alignment with human reviews was reported in \result{All}{alignedpoor}. In cohort order, poor alignment was reported by \pc{LC}{alignedpoor}, \pc{LN}{alignedpoor}, and \pc{CV}{alignedpoor}, and any inaccuracies by \pc{LC}{anyerror}, \pc{LN}{anyerror}, and \pc{CV}{anyerror}. These are respondents' assessments of their reviews, not independently established error rates.

    \begin{figure}[htbp]
    \centering
    \begin{tikzpicture}
    \begin{axis}[
     xbar, width=0.73\linewidth, height=6.1cm,
     xmin=0, xmax=110, xlabel={Responses (\%)},
     label style={font=\footnotesize}, tick label style={font=\footnotesize},
     xtick={0,20,40,60,80,100}, ytick={1,2,3,4},
     yticklabels={{\shortstack[r]{Was the AI-based review poorly\\aligned with human reviews?}},{\shortstack[r]{Any incorrect or\\misleading comments?}},{\shortstack[r]{Minor inaccuracies,\\but mostly correct?}},{\shortstack[r]{Clearly incorrect/misleading\\or irrelevant comments?}}}, y dir=reverse,
     y tick label style={anchor=east,font=\footnotesize,xshift=-2pt},
     bar width=6pt, enlarge y limits=0.17, axis lines*=left,
     legend style={at={(0.5,-0.20)},anchor=north,draw=none,font=\footnotesize},
     legend columns=3,
     xmajorgrids=true, grid style={dashed,gray!40}, point meta=x,
     nodes near coords={\pgfmathprintnumber[fixed,precision=0]{\pgfplotspointmeta}},
     nodes near coords style={font=\scriptsize,xshift=2pt}]
    \addplot[fill=blue!70!black,draw=blue!70!black,bar shift=8.5pt] coordinates {
(50.00000000,1)
(66.66666667,2)
(50.00000000,3)
(16.66666667,4)
};
\addplot[fill=orange!85!black,draw=orange!85!black,bar shift=0pt] coordinates {
(22.85714286,1)
(54.28571429,2)
(42.85714286,3)
(11.42857143,4)
};
\addplot[fill=black!75,draw=black!75,bar shift=-8.5pt] coordinates {
(33.33333333,1)
(70.37037037,2)
(55.55555556,3)
(14.81481481,4)
};
\legend{LifeCLEF 2025 ($n=30$),LifeCLEF 2026 ($n=35$),CVWW 2026 ($n=27$)}
\end{axis}\end{tikzpicture}
\caption{\textbf{Reported limitations and sources of misfit in the AI-based review.} ``Poorly aligned'' combines ratings 1--2. ``Any'' includes all responses except ``No, all comments were accurate.'' Minor and clearly incorrect/misleading or irrelevant categories use the full cohort denominator, not only responses reporting errors. Sample sizes: LifeCLEF 2025 $n=30$, LifeCLEF 2026 $n=35$, CVWW 2026 $n=27$.}
\label{fig:limitations_misfit}
\end{figure}

\noindent\textbf{Actionability, future use, and governance.}
Reported use in camera-ready revision was common: \result{All}{incorporated} incorporated at least some AI feedback, although only \result{All}{full} selected full incorporation, defined by the questionnaire as implementing most or all suggestions (Fig.~\ref{fig:actionability_acceptance}). Incorporation ranged from \pc{CV}{incorporated} in CVWW to \pc{LN}{incorporated} in LifeCLEF 2026. Prior AI self-review was reported in \result{All}{prior}, while \result{All}{internal} would consider it before future submissions. This comparison describes past use and future intention, rather than a measured change in behavior.

Support for future review by both humans and AI was \result{All}{supplement}. Advance notice was preferred in \result{All}{notice}, and explicit consent under either affirmative option was supported in \result{All}{consent}. Consent preferences varied more than notice preferences: support was \pc{LC}{consent}, \pc{LN}{consent}, and \pc{CV}{consent} across the three cohorts. The most common concerns response was ``Minor concerns (e.g., should be used with supervision)'' (\result{All}{supervision}); \result{All}{majorconcerns} reported major concerns. Thus, willingness to use AI feedback coexisted with reservations about its reliability and administration.

    \begin{figure}[htbp]
    \centering
    \begin{tikzpicture}
    \begin{axis}[
     xbar, width=0.73\linewidth, height=7.0cm,
     xmin=0, xmax=110, xlabel={Responses (\%)},
     label style={font=\footnotesize}, tick label style={font=\footnotesize},
     xtick={0,20,40,60,80,100}, ytick={1,2,3,4,5},
     yticklabels={{\shortstack[r]{Included some AI feedback\\in the camera-ready?}},{\shortstack[r]{Future review by\\both humans and AI?}},{\shortstack[r]{Consider AI for internal\\review before submission?}},{\shortstack[r]{Prefer to be told\\in advance?}},{\shortstack[r]{Support explicit\\author consent?}}}, y dir=reverse,
     y tick label style={anchor=east,font=\footnotesize,xshift=-2pt},
     bar width=6pt, enlarge y limits=0.17, axis lines*=left,
     legend style={at={(0.5,-0.20)},anchor=north,draw=none,font=\footnotesize},
     legend columns=3,
     xmajorgrids=true, grid style={dashed,gray!40}, point meta=x,
     nodes near coords={\pgfmathprintnumber[fixed,precision=0]{\pgfplotspointmeta}},
     nodes near coords style={font=\scriptsize,xshift=2pt}]
    \addplot[fill=blue!70!black,draw=blue!70!black,bar shift=8.5pt] coordinates {
(83.33333333,1)
(80.00000000,2)
(96.66666667,3)
(86.66666667,4)
(63.33333333,5)
};
\addplot[fill=orange!85!black,draw=orange!85!black,bar shift=0pt] coordinates {
(100.00000000,1)
(97.14285714,2)
(97.14285714,3)
(91.42857143,4)
(82.85714286,5)
};
\addplot[fill=black!75,draw=black!75,bar shift=-8.5pt] coordinates {
(81.48148148,1)
(100.00000000,2)
(96.29629630,3)
(92.59259259,4)
(92.59259259,5)
};
\legend{LifeCLEF 2025 ($n=30$),LifeCLEF 2026 ($n=35$),CVWW 2026 ($n=27$)}
\end{axis}\end{tikzpicture}
\caption{\textbf{Actionability and conditions of acceptance for AI reviews.} Incorporation combines partial and full use. Supplementary review means ``Yes, by both humans and AI.'' Consent combines both affirmative options, including consent except for anonymized manuscripts. Sample sizes: LifeCLEF 2025 $n=30$, LifeCLEF 2026 $n=35$, CVWW 2026 $n=27$.}
\label{fig:actionability_acceptance}
\end{figure}

\section{Qualitative comments}

Authors often described both useful suggestions and reasons to question the AI review. The four themes below explain what they valued, where they encountered problems, and what they expected from those providing the reviews.

\paragraph{Theme 1: Complementary feedback with selective value.}
AI reviews offered another perspective on clarity, implementation details, and reproducibility. Authors often found these suggestions useful while seeing human reviewers as better placed to assess the scientific contribution. A CVWW respondent put it this way: \textit{``The AI review seems complementary to humans. It is not good for ideas novelty etc, but nice for text clarity and justifications.''} A LifeCLEF 2026 respondent described AI feedback as \textit{``particularly good at spotting structural gaps like missing citations and reproducibility details''}, although some criticisms concerned stylistic choices. Even comments on points already addressed could prompt clearer explanations. One LifeCLEF 2025 author wrote: \textit{``Notwithstanding, we adjusted the text to more directly cover those points.''}

Additional comments were not always helpful. A LifeCLEF 2025 respondent found human feedback more substantive, while a CVWW respondent questioned whether ranking AI against human reviews made sense when they offered different kinds of feedback. Another LifeCLEF 2025 respondent saw little reason for venues to provide AI reviews that authors could obtain themselves. Length could also make useful suggestions harder to find. A LifeCLEF 2026 author wrote: \textit{``Overall human feedback was more pertinent and concise while AI review was wordy and useful comments were drowned in somewhat pertinent feedback.''} The same respondent questioned whether issues missed by human reviewers were necessarily relevant. Authors therefore had to decide which additional comments deserved attention, rather than simply treating more feedback as better feedback.

\paragraph{Theme 2: Standards, severity, and contextual fit.}
Some criticisms seemed reasonable in isolation but inappropriate for the paper or venue. A LifeCLEF 2025 respondent explained: \textit{``The paper was part of a challenge (so more empirical) and it treated it more like a novel method paper.''} Authors also objected to requests for unavailable data or extensive experiments beyond the scope of working notes. LifeCLEF 2026 responses questioned demands for ranking details and inaccessible private-leaderboard scores, and the treatment of requests for clarification as major weaknesses. A CVWW respondent found the review detailed and useful but noted: \textit{``It was not calibrated to CVWW, more to CVPR and similar conferences.''} Another questioned whether concerns about compute comparisons should block acceptance.

Authors did not simply prefer less demanding reviews. A LifeCLEF 2025 respondent was puzzled by requirements beyond the competition objectives, yet wrote: \textit{``However, I believe that being reminded of these three aspects was still quite useful to me.''} Others accepted criticisms of weaknesses they already recognized. What mattered was whether a demand was justified for the paper being reviewed and whether its stated severity matched the problem.

\paragraph{Theme 3: Factual reliability and the limits of evidence checking.}
Authors reported both factual mistakes and failures to notice information in their papers. A LifeCLEF 2025 respondent stated: \textit{``Some of the AI's citations were to non-existing articles''}; others disputed descriptions of ensembling or data availability. LifeCLEF 2026 authors reported incorrect rankings, errors in citation details, and a missed repository hyperlink. One wrote: \textit{``It seemed to confuse citation style with absence of citation.''} CVWW responses described misread equations, overlooked captions, and results called unsupported despite being reported in the paper.

Other disagreements were less straightforward. CVWW respondents differed on whether the AI had considered supplementary material. A LifeCLEF 2026 author acknowledged that some details appeared in linked code but were not fully documented in the manuscript. A CVWW author recognized that a mathematical criticism concerned inputs impossible in practice, but still found it useful because the paper had not adequately explained the practical input range. These accounts distinguish mistakes in the review from gaps in the paper, information supplied elsewhere, and differences of judgment. Calling all such comments hallucinations would obscure what authors needed to check or revise.

\paragraph{Theme 4: Author control and accountable use.}
Authors wanted a say in whether and how their manuscripts were processed by AI. A LifeCLEF 2025 respondent wrote: \textit{``We would have probably consented, but the decision was taken away from us.''} Other responses requested advance notice, clear labeling, and disclosure of review criteria. CVWW authors raised concerns about third-party processing and differences between their privacy expectations and the use of ChatGPT; some preferred local or open-weight systems. These were accounts of participants' expectations, not independent assessments of a provider's privacy guarantees.

Even authors who welcomed AI feedback often wanted human verification. A LifeCLEF 2026 respondent said it should be treated as \textit{``advisory evidence rather than as an authoritative judgment''} and requested information about the model and review process. Others wanted explanations of confidence labels and human checks before reviews were released. There was disagreement about how far these safeguards should extend: one LifeCLEF 2026 respondent considered consent necessary for training but not inference, while a CVWW respondent supported automatic screening for clear rejections. Support for transparency and accountability thus did not amount to agreement on every use of AI or every consent requirement.

\section{Discussion}\label{discussion}

Authors frequently found auxiliary AI reviews useful and acted on their suggestions without granting them the same authority as human peer review. This distinction between practical value and trust is the study's central finding. High reported usefulness and incorporation coexisted with frequent inaccuracies, lower comparative trust, and preferences for supervision. The qualitative accounts explain this combination as selective use: authors could extract worthwhile suggestions while rejecting, qualifying, or checking others.

Perceived complementarity was substantial in both directions. AI reviews often raised issues absent from human reports, and human reviews often raised issues absent from AI feedback. Yet a larger set of comments is not necessarily better feedback. Participants distinguished improvements in presentation and reporting from advances in scientific substance, and some questioned whether comparative scales could adequately represent such different contributions. Additional comments also imposed a filtering burden when their relevance was uncertain. Evaluations of auxiliary review should therefore assess the value and cost of actionable feedback, rather than equate comment volume or non-overlap with quality.

Contextual calibration and factual reliability were related but distinct concerns. Applying inappropriate novelty standards, demanding infeasible experiments, or promoting minor issues to acceptance blockers can reduce a review's usefulness even when its individual statements are plausible. Conversely, an apparently well-calibrated review can misread a caption, overlook an equation, or make an incorrect citation claim. Demanding feedback was sometimes valued, including when it introduced criteria authors had not considered. The relevant distinction is therefore between warranted and unwarranted demands, not simply between favorable and unfavorable reviews.

The two LifeCLEF cohorts illustrate contextual variation: LifeCLEF 2026 showed higher usefulness and lower poor-alignment ratings than LifeCLEF 2025, alongside a more task-specific prompt. This observational contrast is compatible with better calibration but cannot isolate prompt effects because model, respondent composition, timing, and procedures also differed. Across the corpus, reporting and reproducibility checks were useful but remained vulnerable to missed links, citation-style confusion, and ambiguity about material documented outside the manuscript. Even a bounded editorial-checking role requires reliable evidence access and human judgment about relevance.

\bigskip
\noindent\textbf{Implications for peer review policy and practice.}
The findings favor investigating AI as an optional supplementary or pre-submission service, with scientific judgment and editorial responsibility retained by people. Support for human--AI reviewing was broad, but did not establish willingness to replace human reviewers. One respondent's support for automated rejection screening illustrates the range of views rather than evidence that such screening is reliable. The distinction between an advisory comment and a decision-bearing judgment should remain explicit.

Governance preferences were also prominent. Advance notice and consent received broad support, while free text connected manuscript privacy, clear labeling, and human verification to acceptability. Preferences were not uniform: consent support varied across cohorts, and one respondent distinguished inference from training. These results support meaningful author choice but do not identify a universally preferred technical configuration. The four conditions in Box~1 synthesize these findings with published concerns about accountability and confidentiality~\cite{hosseini2023fighting,leung2023best,bhavsar2025policies}. They are proposals for practice, not experimentally validated safeguards.

\begin{mdframed}[nobreak=true,linewidth=0.8pt, innertopmargin=6pt, innerbottommargin=6pt,
                  innerleftmargin=10pt, innerrightmargin=10pt]
\textbf{Box 1: Proposed conditions for auxiliary AI review}
\begin{enumerate}[topsep=3pt,itemsep=1pt,parsep=0pt]
  \item \textbf{Define and label its role.} Identify AI-generated feedback clearly and distinguish advisory suggestions from human reviews and editorial decisions.
  \item \textbf{Explain the process in advance.} Describe the system, review criteria, human checks, and manuscript data handling in terms authors can understand.
  \item \textbf{Provide meaningful author choice.} Seek explicit permission before sending unpublished manuscripts to third-party AI services, with a clear opportunity to decline.
  \item \textbf{Retain human accountability.} Check factual claims, relevance, and severity before releasing feedback; keep scientific judgments and editorial decisions under human responsibility.
\end{enumerate}
\end{mdframed}

\bigskip
\noindent\textbf{Interpretation in relation to prior work.}
This study complements research on AI assistance to reviewers by examining the experience of authors receiving feedback. A randomized study of LLM feedback to reviewers assessed review quality within a conference workflow~\cite{thakkar2026large}; our outcomes instead concern perceived usefulness, revision, and governance. A survey of authors who received AI and human reviews found a preference for pre-submission AI use~\cite{lemberger2026authors}, consistent with the strong interest in internal review here. Our study examines these experiences across three computer-science cohorts and connects reported revision with concrete accounts of error and governance. Self-reported incorporation cannot establish improvement in manuscripts.

Concerns about biased or poorly contextualized judgment have also been raised in conceptual work on AI review~\cite{hosseini2023fighting}. The present accounts illustrate possible consequences of inappropriate standards and severity judgments without identifying their underlying mechanisms. Editors' mixed attitudes toward AI chatbots~\cite{ng2025attitudes} and variation in publisher policies~\cite{bhavsar2025policies} further contextualize authors' conditional acceptance. Together, these findings motivate evaluations that consider usefulness, verification burden, confidentiality, and accountability alongside technical performance.

\bigskip
\noindent\textbf{Ethical lessons.}
The distinction between willingness to use feedback and authorization to process a manuscript is consequential. In LifeCLEF 2025, authors received email notification but had no separate opt-in or opt-out before processing. CVWW 2026 and LifeCLEF 2026 required explicit opt-in before AI processing. Survey participation was a separate decision from manuscript processing. The LifeCLEF 2025 account of willingness to consent coupled with dissatisfaction about lost choice shows why retrospective comfort cannot substitute for prospective authorization. Privacy expectations likewise concerned the actual route and conditions of processing, not only whether authors welcomed the resulting review.

\section{Limitations}\label{limitations}

The samples were small and self-selected, and recruitment centered on authors preparing camera-ready papers. Authors of rejected papers may evaluate both AI and human feedback differently. Opt-in to AI review in CVWW 2026 and LifeCLEF 2026 may also select authors more receptive to AI. Responses are clustered within papers and may also come from the same individuals, including across years. Optional attribution names do not permit a complete count of distinct people. Pooled percentages therefore summarize the observed responses rather than estimate independent author-level prevalence.

The study relies on self-reported usefulness, inaccuracies, and incorporation of feedback. It does not independently assess review correctness, changes in manuscript quality, editorial decisions, or publication outcomes. Disagreement with a reviewer is not necessarily evidence of error, and the questionnaire's use of ``hallucinations'' may have encouraged respondents to group factual mistakes, misinterpretation, and unwanted criticism together. Several accounts also questioned whether unlike forms of feedback could be meaningfully ranked on a single comparative scale. No psychometric validation of these items was conducted.

Cohorts differed in paper type, seniority, prompt, model, timing, and review procedure. These differences preclude causal attribution to prompt design or model capability. All settings were within computer science, including challenge working notes and a regional workshop. Transferability to other disciplines, journal articles, and more selective review environments requires investigation.

The qualitative material consists of brief optional written accounts rather than interviews with opportunities for clarification. The investigators' involvement in organizing the workflows introduces confirmation and social-desirability risks. Extract-level traceability and attention to counterexamples improve transparency but do not provide independent validation. No member checking or external qualitative audit was conducted.

Future studies should recruit across acceptance outcomes and disciplines, document review-generation and consent procedures prospectively, and combine author accounts with blinded independent assessment of feedback and resulting revisions. Such designs could distinguish genuine improvements in scientific reporting from additional work generated by inaccurate or poorly prioritized comments.

\section{Conclusions}\label{conclusions}

Across three computer-science cohorts, authors commonly found auxiliary AI reviews useful and reported using their suggestions, while retaining reservations about trust, factual reliability, contextual fit, and severity. Perceived complementarity with human reviews supported selective use, but additional comments did not necessarily provide additional scientific value. The findings favor a supplementary role with clear labeling, advance disclosure, meaningful author choice, and human accountability. They do not establish that AI can replace expert judgment or that any prompt or model is superior.

\section*{List of abbreviations}
\begin{tabular}{ll}
AI & Artificial intelligence \\
API & Application programming interface \\
CHERRIES & Checklist for Reporting Results of Internet E-Surveys \\
CLEF & Conference and Labs of the Evaluation Forum \\
CMT & Conference Management Toolkit \\
CVWW & Computer Vision Winter Workshop \\
GPT & Generative Pre-trained Transformer \\
LLM & Large language model \\
SRQR & Standards for Reporting Qualitative Research \\
\end{tabular}

\section*{Declarations}

\subsection*{Ethics approval and consent to participate}
Ethics approval was granted by the Ethics Committee of the University of West Bohemia in Pilsen (Decision V13/2026, 11 May 2026; Appendix~\ref{app:ethics}). The LifeCLEF 2025 and CVWW 2026 surveys preceded this approval. 

Questionnaire participation was voluntary. In all three cohorts, email invitations described the research purpose, provided the questionnaire link, and treated submission as consent to participate. Permission for manuscript processing was separate: LifeCLEF 2025 used advance notification without a separate opt-in or opt-out, whereas CVWW 2026 and LifeCLEF 2026 required explicit opt-in before AI processing. Papers without affirmative permission in the opt-in cohorts were not processed. Respondents could choose named acknowledgement; substantive responses and quotations are not attributed to names or papers.

\subsection*{Consent for publication}
Quotations are presented without participant names or paper identifiers. Named acknowledgements are included with permission and are separate from the attribution of substantive survey responses.

\subsection*{Availability of data and materials}
A reproducibility package containing de-identified quantitative data and Python analysis code is archived on Zenodo: \url{https://doi.org/10.5281/zenodo.22745907}~\cite{leblanc2026reviewdata}. These materials reproduce all quantitative summaries, Table~\ref{tab:respondents}, and Figures~\ref{fig:usefulness_value_complementarity}--\ref{fig:actionability_acceptance}. The platform-independent analysis uses Python 3.9 or newer; PDF figure rendering requires pdfLaTeX and PGFPlots. The release is licensed under CC BY 4.0. Raw questionnaire exports and complete free-text responses are not publicly released because of potential participant identifiability. The review prompts and questionnaire are reproduced in Appendices~\ref{app:prompts} and~\ref{app:questionnaire}.

\subsection*{Competing interests}
The authors declare no financial competing interests. Their involvement in designing and organizing the AI-review workflows is described in Methods.

\subsection*{Funding}
This research received no external funding.

\subsection*{Authors' contributions}
CL and LP contributed equally to study design, review prompts, questionnaire administration, analysis and interpretation, and writing and revision. Both authors read and approved the final manuscript. OpenAI Codex (GPT-6 Astra, high reasoning setting) assisted analysis-code preparation and manuscript drafting. Both authors closely supervised its use, verified the outputs, critically revised the manuscript, and take full responsibility for the final work.

\subsection*{Acknowledgements}
We thank the organizing teams of LifeCLEF (Alexis Joly, Lukáš Adam, Stefan Kahl, Klára Janoušková, Hervé Goëau, Kevin Barnard, and Robert Bossy) and CVWW (Václav Hlaváč, Ondřej Chum, Eva Matysková, Petra Ivaničová, and Miroslav Purkrábek), and all authors who completed the questionnaires. In particular, we thank the following authors: Lianping Lu, Heng Yang, Shuo Li, Fang Liu, Puhua Chen, Wenping Ma, Murilo Gustineli, Anthony Miyaguchi, Adrian Cheung, Divyansh Khattak, Andrea Menco Tovar, Jairo Serrano, Juan Carlos Martinez-Santos, Edwin Puertas, Jason Kahei Tam, Jingyin Tan, Aiguo Wang, Gleb Tikhonov, Dmitry Tikhonov, Tim Chopard, Darren Rawlings, Hanna Herasimchyk, Robin Labryga, Tomislav Prusina, Luciano Araujo Dourado Filho, Almir Moreira da Silva Neto, Rodrigo Pereira David, Rodrigo Tripodi Calumby, Volodymyr Sydorskyi, Fernando Gonçalves, Svyatoslav Lanskikh, Grigory Demidov, Danis Dinmuhametov, Andrey Khlopotnukh, Kristian Bogdan, Roman Pakhomov, Chandrasekaran Maruthaiyannan, Charles R. Clark, Jack Etheredge, Nelly Semenova, Janez Perš, Jonas Serych, Grega Šuštar, Matej Kristan, Jiří Vyskočil, Nikola Marić, Tomas Zelezny, Lev Kokovkin, Kritika Benjwal, Desmond Oh, Rashmi Banthia, Sree Krishna S, Renuka Shanmugam, Dhruv Pai Dukle, Venkata Pushpak Teja Menta, stephen foy, Antoine Masquelier, Mathieu D. Weill, Musa Tur Farazi, Hanif Noer Rofiq, Seifeldin Elkerdany, Gilles Colling, Pin-Hao Shen, Mohamed ElBassat, Edidiong-Abasi Ita Anwanane, Dan Zimmerman, and Mihai Radu Radulescu.

\subsection*{Additional files}
\textbf{Additional file 1 (PDF): CHERRIES checklist.} The completed online-survey reporting checklist, also included as Appendix~\ref{app:cherries}.

\textbf{Additional file 2 (PDF): SRQR checklist.} The completed qualitative-reporting checklist, also included as Appendix~\ref{app:srqr}.

\bigskip
\begingroup
\let\originalthebibliography\thebibliography
\renewcommand{\thebibliography}[1]{\originalthebibliography{#1}\raggedright}
\bibliography{bibliography}
\endgroup

\clearpage
\singlespacing
\appendix
\section{AI Prompts}
\label{app:prompts}

{\raggedright The following prompts are reproduced in full: LifeCLEF 2025, ChatGPT-4o (\texttt{gpt-4o-2024-11-20}); LifeCLEF 2026, ChatGPT 5.5 (\texttt{gpt-5.5-2026-04-23}); and CVWW 2026, ChatGPT-5.2 (\texttt{gpt-5.2-2025-12-11}). The LifeCLEF 2025 follow-up instruction is included. Prompt wording and citation syntax are unchanged; long lines wrap for readability.\par}

\appqsection{Prompt A; LifeCLEF 2025 Review Prompt}

\begin{promptbox}
\begin{Verbatim}[fontsize=\footnotesize,breaklines,breakanywhere]
You are acting as a peer reviewer for a scientific workshop in computer science: LifeCLEF 2025, part of CLEF.
The workshop focuses on biodiversity informatics challenges involving machine learning, computer vision, and related techniques. The emphasis of the submissions is on reproducibility, careful experimental evaluation, and thoughtful analysis, rather than purely on novelty.
Please carefully read the following paper submission and write a professional, constructive review. Your review should include the following sections:
1. Summary:
    * Briefly summarize the task, methods, datasets, and key findings.
    * State clearly what problem the authors are addressing, which LifeCLEF challenge it pertains to, and what their main contributions are.
2. Strengths: List the strong aspects of the paper, such as:
    * Reproducibility (e.g., availability of code, data, clear methodology)
    * Careful experimental design
    * Well-performed ablation studies or error analyses
    * Insightful discussions of results
    * Clarity of writing and presentation
3. Weaknesses / Areas for Improvement: Identify any weaknesses or limitations, such as:
    * Missing details that would prevent reproduction
    * Lack of ablations or sensitivity analyses
    * Incomplete or unclear description of the method
    * Insufficient discussion or interpretation of the results
    * Missing comparison to appropriate baselines
4. Detailed Comments:
    * Provide actionable, constructive feedback that the authors can use to improve their paper.
    * You may point out specific sections, figures, or tables that need clarification, expansion, or correction.
    * Comment on both scientific and presentational aspects.
5. Overall Evaluation: Please provide your overall recommendation, choosing one of:
    * Strong Accept
    * Accept
    * Weak Accept
    * Borderline
    * Weak Reject
    * Reject
    * Strong Reject

Important Reviewing Guidelines:
* Focus on scientific rigor, reproducibility, and clarity rather than novelty alone.
* Do not hallucinate or infer information not present in the submission.
* Be neutral, unbiased, and professional.
* If some required information is missing, state it explicitly.
\end{Verbatim}
\end{promptbox}

\noindent Follow-up instruction used for LifeCLEF 2025:
\begin{promptbox}
\begin{Verbatim}[fontsize=\footnotesize,breaklines,breakanywhere]
Act more like a Reviewer 2.
\end{Verbatim}
\end{promptbox}

\appqsection{Prompt B; LifeCLEF 2026 Review Prompt}
\begin{promptbox}
\begin{Verbatim}
You are reviewing a LifeCLEF 2026 working notes paper describing a competition solution.

These papers are not meant to be full archival conference papers. They are short working notes describing systems submitted to LifeCLEF tasks. 
Review them accordingly: be fair, direct, and practical. The main question is whether the paper clearly explains a valid, useful, and reproducible solution for the relevant LifeCLEF task.
Do not over-focus on novelty. A simple method can be a good working notes paper if it is well explained, follows the challenge rules, reports results honestly, and helps other participants understand what works.

## What to look for

Read the whole paper, including tables, figures, captions, references, and any appendix or supplement if present.

Focus on:

1. Task-specific grounding
   - Is the LifeCLEF 2026 task, track, phase, or dataset clearly identified?
   - Does the paper make clear which official competition setting it addresses?
   - Does it describe the expected input data and output format for that task?
   - Does it explain how the submitted system maps to the task requirements, such as classification labels, ranked predictions, detections, localizations, species identifications, extracted entities/relations, or other official outputs?

2. Protocol and results
   - Are the evaluation settings clear?
   - Are the metrics named, explained, or properly referenced?
   - Are train/validation/test, public/private leaderboard, development/test, or official phase settings clearly distinguished?
   - Are local validation results separated from official leaderboard results?
   - Are the achieved rank, team name, and leaderboard or phase clearly reported?
   - If a rank or score is claimed, is there enough information to trust it?

3. Method clarity
   - Can another participant roughly reproduce the approach?
   - Are the model, inputs, preprocessing, data filtering, augmentation, training, inference, thresholds, prompts, postprocessing, and ensembling steps described?
   - Are important implementation details missing?
   - Is it clear which components were trained by the authors and which were reused from external models, pretrained checkpoints, APIs, or prior work?

4. Code and reproducibility
   - Is code available, or does the paper clearly state whether it will be released?
   - Are dependencies, checkpoints, prompts, configuration details, hardware, training time, inference time, or runtime constraints reported when relevant?
   - Is there enough information for another team to rebuild the system or understand why it performed as reported?

5. Figures, images, and tables
   - Are figures readable and useful?
   - Are images of sufficient quality?
   - Are captions informative?
   - Are tables clear, consistently formatted, and connected to the text?
   - Do tables specify the metric, split, phase, or leaderboard for each reported result?

6. Language and presentation
   - Is the paper understandable and professionally written?
   - Are there grammar, terminology, formatting, or organization issues that affect clarity?
   - Are claims supported by the described method or results?
   - Is there generic filler text, vague wording, or unsupported interpretation that should be removed?

7. AI-related artifacts
   - Are there visible AI-generated artifacts, such as leftover prompt text, “as an AI language model,” generic generated-looking filler, fabricated-sounding claims, or references to experiments, tools, datasets, or results that are not actually described?
   - Do any sections appear internally inconsistent or disconnected from the actual submitted system?

8. References and required citations
   - Does the paper cite the general LifeCLEF 2026 lab overview paper?
   - Does the paper cite the relevant task overview paper?
   - Does the paper cite the relevant datasets, benchmarks, pretrained models, libraries, external APIs, or prior systems used?
   - Are there suspicious, incomplete, or probably fake references?
   - Are there citations in the text that are missing from the bibliography?
   - Are there bibliography entries that are never cited?

## Required citations

The paper should cite the LifeCLEF 2026 lab overview paper and the relevant competition overview paper for the task it addresses.

### General LifeCLEF citation

All LifeCLEF 2026 working notes papers should cite:

@inproceedings{lifeclef2026,   title={Overview of {LifeCLEF} 2026: AI Challenges for Biodiversity Understanding and Ecosystem Management},   author={Picek, Lukas and Adam, Luk{\'a}{\v{s}} and Kahl, Stefan and Bossy, Robert and Chrobak, Laura and Go{\"e}au, Herv{\'e} and Papafitsoros, Kostas and Klinck, Holger and Vellinga, Willem-Pier and Planqu{\'e}, Robert and Denton, Tom and Barnard, Kevin and N{\'e}dellec, Claire and Del{\'e}ger, Louise and Courtin, Marine and Martellucci, Giulio and Moummad, Ilyass and Vinatier, Fabrice and Bonnet, Pierre and Joly, Alexis},   booktitle={International Conference of the Cross-Language Evaluation Forum for European Languages (CLEF)},   year={2026},   organization={Springer} } 

### Task-specific overview citations

The paper should also cite the relevant task overview paper.

For AnimalCLEF submissions, the expected citation is:

bibtex @inproceedings{animalclef2026overview,   title={Overview of {AnimalCLEF} 2026: Discovery and Re-identification of Individual Animals},   author={Adam, Luk\'a\v{s} and Picek, Lukas and Papafitsoros, Kostas and Williams, Dean A. and Biffi, Daniella},   booktitle={Working Notes of CLEF 2026 - Conference and Labs of the Evaluation Forum},   year={2026} } 

For FathomNetCLEF submissions, the expected citation is:

bibtex @inproceedings{fathomnet026overview,   title={Overview of {FathomNetCLEF} 2026: Positive-Unlabeled Object Detection in Marine Images},   author={Chrobak, Laura and Barnard, Kevin},   booktitle={Working Notes of CLEF 2026 - Conference and Labs of the Evaluation Forum},   year={2026} } 

For BirdCLEF submissions, the expected citation is:

bibtex @inproceedings{birdclef2026overview,   title={Overview of {BirdCLEF+} 2026: Acoustic Species Identification in the {Pantanal, South America}},   author={     Kahl, Stefan and     Denton, Tom and     Sugai, Larissa and     Piatti, Liliana and     Arruda Moreno, Wener Hugo and     Barbosa, Mariana Motti and     Eibl, Maximilian and     Fieker, Carolline Zatta and     Garcia, Carolina Martins and     Hokama Sousa, Daiene Louveira and     J{\'u}nior, Jo{\~a}o Em{\'i}lio de Almeida and     Kridler, Ryan Christopher and     Lasseck, Mario and     Melo, Alyson Vieira de and     M{\"u}ller, Henning and     Neves, Matheus de Oliveira and     Reis, Matheus Gon{\c{c}}alves dos and     Sampaio Teles, Lucas Korzune and     Schuchmann, Karl-L. and     Sugai, Jos{\'e} Luiz Massao Moreira and     Azevedo, Kirk Thiago Pedroso and     Lopes, Priscila do Nascimento and     Marques, Marinez Isaac and     Klinck, Holger and     Glotin, Herv{\'e} and     Go{\"e}au, Herv{\'e} and     Vellinga, Willem-Pier and     Planqu{\'e}, Robert and     Joly, Alexis   },   booktitle={Working Notes of CLEF 2026 - Conference and Labs of the Evaluation Forum},   year={2026} } 

For PestCLEF submissions, the expected citation is:

bibtex @inproceedings{pestclef2026overview,   title={Overview of {PestCLEF} 2026: Knowledge Graph Extraction from Plant Health Literature},   author={Bossy, Robert and Courtin, Marine and Del{\'e}ger, Louise and N{\'e}dellec, Claire},   booktitle={Working Notes of CLEF 2026 - Conference and Labs of the Evaluation Forum},   year={2026} } 

For PlantCLEF submissions, the expected citation is:

bibtex @inproceedings{plantclef2026,   author = {Giulio Martellucci and Ilyass Moummad and Herv{'e} Go{"e}au and Pierre Bonnet and Fabrice Vinatier and Alexis Joly},   title = {Overview of {PlantCLEF} 2026: Identify Multi-species Plants in Images of Vegetation Plots},   booktitle = {Working Notes of CLEF 2026 - Conference and Labs of the Evaluation Forum},   year={2026} } 

Also check whether the paper cites relevant dataset papers, benchmark papers, software, pretrained models, external APIs, or prior methods used in the solution.

Accept small formatting differences if the title, authors, venue, and year are clearly correct. If the LifeCLEF overview paper, the relevant task overview paper, or important dataset/model citations are missing, mention this as a weakness.

## Hallucinated or suspicious references

Be careful here. Do not accuse a paper of hallucinating references unless there is strong evidence.
Search for existence on Google Scholar.

Use these labels:
- Looks valid: the reference is internally consistent and normal-looking.
- Suspicious: the title, venue, year, or author list looks odd, incomplete, or inconsistent.
- Likely hallucinated: the reference appears impossible, internally contradictory, or cannot be found when exact-title search is available.
- Cited but missing: cited in the text but absent from the bibliography.
- Bibliography only: listed in the bibliography but not cited in the paper.
- Duplicate or inconsistent: repeated or conflicting entries.

If you cannot check externally, say that manual verification is needed.

## Evidence anchors

For every factual criticism or praise, point to where it comes from.

Use anchors such as:
- (Sec. 2)
- (Table 1)
- (Fig. 3)
- (Paragraph starting “We train...”)
- (Section “Experiments”)
- (Reference list)

If the paper does not support a point, write: No direct evidence found in the manuscript.
Do not invent evidence.

## Output format

Use exactly these headings:
- Summary of the report
- Strengths
- Weaknesses
- Suggestions for Improvement
- Recommendation: Reject / Weak Reject / Weak Accept / Accept
- Justification

Keep the review around 400–800 words. Be concise. The tone should sound like a careful human reviewer, not a template.

## Summary of the report

Write one short paragraph, at most 120 words.

Summarize:
- what LifeCLEF task or setting the paper addresses,
- what the paper tries to do,
- the main method,
- the main reported result.

Keep this neutral. Do not include recommendation language here.

## Strengths

Give 3–5 bullets.

Each bullet should have:

- a short bold title,
- one or two sentences,
- evidence anchors.

Example:

- Clear task grounding: The paper identifies the AnimalCLEF task and describes the submitted system. (Sec. 1) It also states the official evaluation metric and leaderboard setting. (Sec. 4)

Good strengths might include:

- clear task grounding,
- clear competition setting,
- strong official leaderboard result,
- reproducible pipeline,
- useful ablation,
- useful error or failure analysis,
- clear data preprocessing,
- clear use of external models or pretrained checkpoints,
- available code,
- clear reporting of compute or runtime,
- readable figures and tables,
- well-written short report.

## Weaknesses

Give 3–6 bullets.

Each weakness title must start with one of:
- [Critical]
- [Major]
- [Minor]

Use this meaning:
- [Critical]: could invalidate the submission or make the result impossible to trust.
- [Major]: important problem, but likely fixable.
- [Minor]: clarity, writing, citation, formatting, or presentation issue.

Each weakness should include:
- what is wrong,
- why it matters,
- the evidence,
- whether it blocks acceptance.

Example:
- [Major] Unclear leaderboard setting: The paper reports an official score and rank but does not say whether these refer to the public leaderboard, private leaderboard, final phase, or another evaluation phase. (Table 1) This makes the reported performance difficult to interpret. Decision impact: Blocker.

Include a reference-related weakness if the required LifeCLEF / task / dataset citation is missing, wrong, or if suspicious references appear.

Include a reproducibility-related weakness if the system cannot be roughly reproduced from the description.

Include a figure/table-related weakness only if the issue affects readability, interpretation, or trust in the results.

Include an AI-artifact weakness if visible generated-text artifacts, unsupported claims, or fabricated-sounding content materially affect the paper.

Only include a math/notation weakness if notation or equations create a real problem.

## Suggestions for Improvement

Give one suggestion for each major or critical weakness.

Use the same title as the weakness, but start it with Fix:.

Be concrete. Say exactly what should be added or changed.

Each suggestion should mention:

- what to add,
- where to add it,
- how a reader would know the problem is fixed.

Example:

- Fix: Unclear leaderboard setting: Add a small table in the experiments section listing each reported score, the split or phase, metric, leaderboard source, team name, and whether the result is local or official. The issue is fixed if every number in the paper can be traced to one evaluation setting.

## Recommendation: Reject / Weak Reject / Weak Accept / Accept

Give one label only.

Use this calibration:

- Accept: clear, valid, useful LifeCLEF working notes paper with no blocking issues.
- Weak Accept: good enough for the workshop, but still needs some cleanup.
- Weak Reject: relevant and possibly useful, but important issues must be fixed first.
- Reject: unverifiable claims, severe protocol problems, missing or invalid submission description, or a system description that is too vague to review.

Format:

Recommendation: <label> — <one-sentence reason with evidence anchor>.

## Justification

Write one short paragraph, 4–6 sentences.

Mention:

- the main reason to accept,
- the main reason to hesitate,
- whether the paper is useful to LifeCLEF participants,
- what would most improve it,
- confidence level: High, Medium, or Low.

Every factual sentence should have an evidence anchor.
\end{Verbatim}
\end{promptbox}

\appqsection{Prompt C; CVWW 2026 Review Prompt}

\begin{promptbox}
\begin{Verbatim}[fontsize=\footnotesize,breaklines,breakanywhere]
## [Role]
You are an expert CVPR / ICCV / NeurIPS reviewer (computer vision + ML) with strong experience in:
- vision benchmarks and evaluation protocols (splits, metrics, leakage, statistical reporting)
- model/system design (architectures, training details, compute/memory/latency trade-offs)
- rigorous ablations, fair comparisons, reproducibility, and failure analysis
- careful novelty/positioning vs. prior vision literature

Your job: produce a text-only, structured review and subsequently propose recommendation.
Important: be candid and rigorous. This is for a self-review to improve the paper before submission.

## [What you will receive]
- Full anonymous manuscript text
- (Optional) Supplementary text

## [Evidence Anchoring Rules]
Every factual claim you make MUST include an evidence anchor at the end of the sentence.
Allowed evidence anchors:
- If numbered: (Sec. 3.2), (Eq. 5), (Fig. 2), (Table 4), (Alg. 1), (Appx. B)
- If NOT numbered: (Section ``Experiments''), (Figure caption starting ``...''), (Paragraph starting ``...''), (Equation defining ``...'')
If you cannot anchor: write ``No direct evidence found in the manuscript.''

Do NOT cite external sources unless they appear in the manuscript's reference list (and you can verify they appear there).

## [Review Process — Do internally before writing]
1) Comprehension pass:
- Identify problem, motivation, and setting (task, inputs/outputs, supervision)
- Extract claimed contributions (aim for 3–6 bullet-level items internally)
- Identify method components and training/inference pipeline
- Locate experiments: datasets, metrics, baselines, evaluation protocol, qualitative results

2) Critical pass:
- Check internal consistency: do the claims match the shown evidence?
- Check methodological clarity: could a reader reproduce it?
- Check experimental fairness: same data, same compute budgets if claimed, correct protocols
- Check statistical rigor when applicable (variance, seeds, CIs, significance claims)
- Check qualitative evidence: are visuals meaningful or cherry-picked? are failure cases shown?

3) Conference-specific scrutiny:
- Novelty/positioning vs. closest prior work in vision
- Practicality: compute, memory, inference speed, scaling behavior
- Robustness: domain shift, corruptions, hyperparameter sensitivity, ablations
- Ethics/societal impact if relevant (privacy, surveillance, dataset bias, misuse)

## [Output Constraints]
Use EXACTLY these headings in this order (no additions, no omissions):
- Summary of the paper
- Strengths
- Weaknesses
- Suggestions for Improvement
- Recommendation: Reject / Weak Reject / Weak Accept / Accept
- Justification

## [Concision Constraints (MANDATORY, paper-dependent)]
Your review must be concise and high-signal, not exhaustive.

Global limits:
- Total length: 700–1200 words (aim for the low end unless the paper is unusually complex).
- Prefer fewer, higher-impact items over many small ones.
- No sentence longer than 30 words.
- Avoid filler/meta (``overall'', ``in general'', ``it is worth noting'', ``clearly'') unless followed by a concrete anchored detail.
- Do not restate the method more than once outside ``Summary of the paper''.
- If a point is not decisive for correctness, novelty, experimental validity, or reproducibility, omit it.

Item budgeting (flexible, not ``exactly''):
- Strength items: typically 3–5.
- Weakness items: typically 4–6.
- Suggestions: 4–6 items, prioritizing the highest-ROI fixes.
Choose counts based on what the manuscript supports while staying within the word limit.

## Summary of the paper
**Reasoning Process:**
- Extract the core problem statement (what gap does this address?)
- Summarize the proposed method/approach (how does it work?)
- Identify key contributions (what's novel?)
- State main experimental results (what was achieved?)

**Output Requirements:**
- Concisely and neutrally restate problem, method, contributions, and results (≤150 words)
- Avoid subjective judgments or decision-like language
- Focus on factual content only
- No evidence anchors.

## Strengths
- Provide 3–6 items (prioritize the most important).
- Every sentence ends with an evidence anchor (or ``No direct evidence...'').
Format for each item:
- **Strength title (<=7 words)**: Sentence 1: what is strong + where it appears. (evidence anchor). Sentence 2: concrete supporting detail. (evidence anchor). Sentence 3 (optional): why it matters. (evidence anchor)

Try to cover, when evidence exists:
- clarity of problem formulation/motivation
- technical innovation/soundness
- quality of qualitative results/visualizations (CVPR-specific)
- experiment breadth and fairness
- ablations, robustness, and analysis
- reproducibility details (hyperparameters, compute, code/data availability)
- practical trade-offs (params/FLOPs/latency/memory) if relevant

## Weaknesses
Provide as many items as the manuscript supports (≥3 encouraged; prioritize acceptance blockers).
Same bullet structure as Strengths: [Severity] Title + 4–6 evidence-anchored sub-points.
Each sub-point must be a single sentence ending with an evidence anchor (or ``No direct evidence found in the manuscript.'').
Do not speculate beyond the manuscript; if uncertain, state uncertainty explicitly and anchor it.

Severity requirement (mandatory):
Each weakness title must start with [Critical] / [Major] / [Minor].
- [Critical] = threatens correctness, invalidates evaluation, or demonstrates leakage/unfair protocol with evidence.
- [Major] = reduces reproducibility or confidence, but plausibly fixable with added details/experiments.
- [Minor] = clarity or presentation; camera-ready fixes.

Decision-impact requirement (mandatory):
End each weakness item with one final sub-point:
- Decision impact: Blocker / Not a blocker. (evidence anchor or ``No direct evidence found in the manuscript.'')

Include one item titled Mathematical and notational clarity/correctness only if you find issues that are [Major] or [Critical]. In that case, provide specific evidence-anchored sub-points referencing:
- equation definitions, symbol reuse, missing assumptions, unclear derivation steps,
- mismatch between text and equations, ambiguous objectives, or undefined variables.

Try to cover, when evidence exists:
- novelty/positioning gaps vs. closest prior work (only using manuscript references; no external citations)
- missing baselines or unfair comparisons (only claim unfairness if a protocol mismatch is evidenced)
- insufficient protocol detail (splits, preprocessing, training schedule, augmentation, test-time tricks)
- missing variance reporting (seeds, std, CIs), especially if gains are small
- limited generalization (few datasets, narrow domain, no cross-dataset test)
- lack of failure cases / negative results / limitations discussion
- compute and resource reporting gaps; scaling behavior not analyzed
- ethical concerns or dataset bias concerns (if relevant)

## Suggestions for Improvement
Number of suggestion ITEMS does not have to equal the number of Weakness items.
Each suggestion item must:
- have the same **Title** as the corresponding weakness, prefixed with ``Fix: ...''
- contain the SAME number of sub-points as that weakness (one-to-one)

Each sub-point must include:
- an actionable step (exact experiment / rewrite / analysis to add)
- a verifiable success criterion (what result/clarity would confirm it)
- reproducibility details (what to report: seeds, splits, hyperparams range, compute, etc.)
Each sub-point must end with an evidence anchor if it refers to existing content; otherwise specify what should be added and where (e.g., ``Add Table X in Sec. 'Experiments' ...'').
Be concrete enough that another researcher could implement it without ambiguity.

## Recommendation:
- Output ONE label [Reject, Weak Reject, Weak Accept, Accept] per venue tier (CORE A*, CORE A, CORE B, CORE C), then ONE sentence stating the decisive reason for that tier, ending with an evidence anchor (or ``No direct evidence found in the manuscript.'').
- Format each line as: CORE <tier>: <label> — <one decisive sentence>. (...)

## Justification
Write 1–2 short paragraphs (4–7 sentences total).
Include:
- the top 1–2 strengths and why they matter,
- the top 1–2 weaknesses/blockers and why they matter,
- what changes would most likely improve the recommendation,
- confidence level (High/Med/Low) and one reason for it.
Every sentence must end with an evidence anchor or ``No direct evidence found in the manuscript.''

## Final self-check (do internally)
- Every sentence has an evidence anchor or ``No direct evidence found...''
- ≥3 strengths and ≥3 weaknesses if manuscript supports it
- Weaknesses include the required math/notation item
- Suggestions are not one-to-one with weaknesses, with matched sub-point counts
- Tone is rigorous, polite, constructive
- Length roughly 500–1000 words (adjust to manuscript complexity)
\end{Verbatim}
\end{promptbox}

\section{Questionnaire Used in the Study}
\label{app:questionnaire}

{\raggedright The questionnaires used equivalent substantive items with minor venue-specific terminology adaptations (``AI''/``ChatGPT'' and ``working note''/``manuscript''). Items are grouped below by topic; the optional name field was present in CVWW 2026 and LifeCLEF 2026 only.\par}

\begingroup
\tcbset{before skip=3pt,after skip=3pt}
\let\originalquestion\question
\renewcommand{\question}[1]{\Needspace{6\baselineskip}\originalquestion{#1}}
\let\originalappqsection\appqsection
\renewcommand{\appqsection}[1]{\Needspace{8\baselineskip}\originalappqsection{#1}}

\appqsection{1. Identification and Background}

\question{Paper ID \textit{[as appears in attached file / Easychair]}}
\answertype{\textit{~~Free text}}

\question{Do you wish to be credited in potential future publication for answering the survey?}
\inlineanswers{Yes / No}

\question{Name to be credited.}
\answertype{\textit{~~Free text}}

\question{How experienced are you in the field relevant to your submission?}
\possanswers{
\begin{itemize}
    \item Undergraduate (studying for bachelor degree)
    \item Graduate (studying for master degree)
    \item PhD student
    \item Postdoc
    \item Faculty position / Advisor with a PhD
    \item None of the above
\end{itemize}
}

\question{Did you use AI or any similar generative AI tool to assist writing or reviewing the working note you submitted?}
\possanswers{
\begin{itemize}
    \item No
    \item Yes, for writing
    \item Yes, for self-reviewing (i.e., before submitting the paper)
    \item Yes, for both
\end{itemize}
}

\appqsection{2. Usefulness of the Reviews}

\question{Overall, was the AI-generated review useful?}
\inlineanswers{Yes / No}

\question{Overall, were the human reviews useful?}
\inlineanswers{Yes \quad / \quad Some YES, some NO \quad / \quad No}

\question{Compared to human review(s), the AI review was:}
\fivepointscale{Way less useful}{Way more useful}

\question{How aligned was the AI review with the human review(s)?}
\fivepointscale{Way less aligned}{Very aligned}

\appqsection{3. Comparison Between AI and Human Reviews}

\question{Did the AI review identify issues that human review(s) did not mention?}
\inlineanswers{Yes / No}

\question{Did the human review(s) identify issues that the AI review did not mention?}
\inlineanswers{Yes / No}

\question{Who was giving clearer suggestions (i.e., whose suggestions could you use to improve your paper)?}
\inlineanswers{The human(s) \quad / \quad Both were pretty much equal \quad / \quad AI}

\question{How much did you trust the AI review compared to the human review(s)?}
\fivepointscale{Way less}{Way more}

\appqsection{4. Impact on Revision}

\question{Did you incorporate changes suggested by AI into your camera-ready version?}
\possanswers{
\begin{itemize}
    \item Yes, fully --- I implemented most or all of AI's suggestions.
    \item Partially --- I used some of AI's feedback.
    \item No --- I decided not to use AI's suggestions.
\end{itemize}
}

\appqsection{5. Incorrect or Misleading Comments}

\question{Did the AI review include any incorrect or misleading comments (also named ``hallucinations'')?}
\possanswers{
\begin{itemize}
    \item No, all comments were accurate.
    \item Minor inaccuracies, but mostly correct.
    \item Some comments were clearly incorrect or misleading.
    \item Many comments were incorrect or irrelevant.
\end{itemize}
}

\question{If YES, please briefly reference any example of ``hallucination''.}
\answertype{\textit{Free text}}

\appqsection{6. Future Use of AI-Based Review}

\question{In the future, would you like your paper to be reviewed by AI-based reviewers?}
\inlineanswers{Yes, by both humans and AI \quad / \quad No, only by humans \quad / \quad Yes, only by AI}

\question{Would you prefer to be told in advance that your paper will (also) be reviewed by AI?}
\inlineanswers{Yes / No}

\question{Would you consider using AI feedback as an internal review tool before future submissions?}
\inlineanswers{Yes / No}

\appqsection{7. Concerns and Consent}

\question{Do you have any concerns about using AI in peer review?}
\possanswers{
\begin{itemize}
    \item No
    \item Minor concerns (e.g., should be used with supervision).
    \item Major concerns (e.g., should not be used).
    \item Unsure
\end{itemize}
}

\question{Were you comfortable that your manuscript text was input to AI?}
\fivepointscale{Not comfortable at all}{Very comfortable}

\question{Should authors be required to give explicit consent before their submissions are sent to a third-party chatbot for review?}
\possanswers{
\begin{itemize}
    \item Yes, no matter what
    \item Yes, except if the manuscript is anonymized
    \item No
    \item I'm not sure / I can't tell
\end{itemize}
}

\appqsection{8. Final Comments}

\question{Is there anything else you would like to share regarding this experiment or your experience?}
\answertype{Free text}
\endgroup

\section{CHERRIES Checklist}
\label{app:cherries}

\noindent Online-survey reporting follows Eysenbach~\cite{eysenbach2004cherries}.

\begingroup
\footnotesize
\renewcommand{\arraystretch}{1.65}
\begin{longtable}{@{}>{\bfseries\raggedright\arraybackslash}p{2.5cm}
                  >{\raggedright\arraybackslash}p{10.5cm}@{}}
\toprule
Checklist item & Study report \\
\midrule
\endhead
\midrule
\multicolumn{2}{r}{\footnotesize\textit{(continued on next page)}}\\
\endfoot
\bottomrule
\endlastfoot
\multicolumn{2}{@{}l}{\bfseries\small Design} \\*[4pt]
Describe survey design & Three related post-review Google Forms surveys of authors receiving one auxiliary AI review alongside at least two human reviews. Participation was voluntary; recruitment focused on camera-ready authors. In cohort order (LifeCLEF 2025, LifeCLEF 2026, CVWW 2026): 31, 34, and 26 physical submissions; 30, 34, and 26 completed forms; 30, 35, and 27 paper-linked quantitative responses; 20, 31, and 20 represented papers. \\[6pt]
Sampling frame & Invitations concerned 89, 82, and 70 author--paper instances. An individual could be invited for multiple papers. LifeCLEF 2026 had 53 accepted papers with 111 author--paper instances; 36 papers opted in to AI review, yielding the 82 eligible survey invitations. The survey frame was restricted to authors receiving AI feedback. \\[6pt]
\midrule
\multicolumn{2}{@{}l}{\bfseries\small Ethics and consent} \\*[4pt]
Ethics approval & Ethics Committee of the University of West Bohemia in Pilsen, Decision V13/2026, 11 May 2026 (Appendix~\ref{app:ethics}). The LifeCLEF 2025 and CVWW surveys preceded approval. \\[6pt]
Informed consent & Email invitations explained the research purpose and linked to the questionnaire; submission constituted consent to voluntary survey participation. Manuscript-processing permission was separate: LifeCLEF 2025 used advance notification without a separate opt-in/opt-out; CVWW and LifeCLEF 2026 required affirmative opt-in before AI processing. Papers without opt-in in these cohorts were not processed. \\[6pt]
Data protection & Raw exports contain paper identifiers and optional names in CVWW and LifeCLEF 2026. For LifeCLEF 2025, requested attribution was resolved at paper level from conference records. Names and paper identifiers are excluded from the analytical free-text table; quotations identify only cohorts. Public quantitative data exclude names, paper identifiers, timestamps, attribution choices, and free text. \\[6pt]
\midrule
\multicolumn{2}{@{}l}{\bfseries\small Development and recruitment} \\*[4pt]
Development and testing & The questionnaire was developed for this pilot. No external pilot study or formal psychometric validation was performed. \\[6pt]
Open vs. closed survey & Eligible authors received direct email invitations with a Google Forms link. There was no public recruitment or advertising and no login requirement. Access was by link rather than authenticated individual accounts. \\[6pt]
Contact mode & Email invitations in all three cohorts, after receipt of AI and human reviews. \\[6pt]
Advertising & No public advertising. Recruitment was within the participating review workflows. \\[6pt]
\midrule
\multicolumn{2}{@{}l}{\bfseries\small Survey administration} \\*[4pt]
Web/E-mail & Three Google Forms instances, one per cohort; responses exported as CSV. \\[6pt]
Context & Questions concerned authors' review experiences and camera-ready revision. The investigators also organized the workflows, which may have influenced responses. \\[6pt]
Mandatory / voluntary & Participation was voluntary. Core items were required; substantive free text and attribution fields were optional where applicable. All 90 nonblank forms had complete core responses. \\[6pt]
Incentives & Optional acknowledgement was offered; no financial or material incentive. \\[6pt]
Time/Date & Observed submission dates: LifeCLEF 2025, 8 December 2025--1 February 2026; LifeCLEF 2026, 13 July--22 August 2026; CVWW, 13 January--10 February 2026. \\[6pt]
Randomization & Fixed item order in all three cohorts. \\[6pt]
Adaptive questioning & Conditional display where appropriate, including optional inaccuracy examples and attribution fields. \\[6pt]
Number of items & Twenty-two questions, including two optional substantive free-text items; CVWW and LifeCLEF 2026 also had an optional name field. Substantive items were equivalent across cohorts (Appendix~\ref{app:questionnaire}). \\[6pt]
Number of screens & Seven screens in each questionnaire. \\[6pt]
Completeness check & Core items were required. One entirely blank LifeCLEF 2025 submission was excluded from substantive analysis. \\[6pt]
Review step & No separate response-review step before final submission. \\[6pt]
\midrule
\multicolumn{2}{@{}l}{\bfseries\small Response rates} \\*[4pt]
Unique site visitors & No visitor-level data were collected for analysis. Questionnaire counts do not establish numbers of unique physical people. \\[6pt]
View rate & Not calculated; questionnaire-page view counts were not collected. \\[6pt]
Participation rate & Operational participation uses author--paper invitation instances: LifeCLEF 2025, \participation{LC}; LifeCLEF 2026, \participation{LN}; CVWW, \participation{CV}; pooled, \participation{All}. Each multi-paper form represents two separately invited instances. The blank LifeCLEF 2025 submission counts as participation but not substantive analysis. These are not unique-person response rates. \\[6pt]
Completion rate & Physical-form completion was 90/91 (99\%); starts and abandonments were not measured. Nonempty inaccuracy examples: 10/30 (33\%), 8/34 (24\%), 10/26 (38\%); final comments: 13/30 (43\%), 15/34 (44\%), 8/26 (31\%). Attribution was requested on 24, 24, and 8 forms. \\[6pt]
\midrule
\multicolumn{2}{@{}l}{\bfseries\small Multiple entries} \\*[4pt]
Cookies / IP check & No cookie- or IP-based duplicate exclusion was performed. \\[6pt]
Log file analysis & No server-log data were analyzed. \\[6pt]
Registration & No login required. Valid separate questionnaires were retained, including responses concerning the same paper or from individuals contributing to multiple papers. One CVWW form and one LifeCLEF 2026 form explicitly covered two papers; each was expanded for paper-linked quantitative analysis, with its original free text considered once. \\[6pt]
\midrule
\multicolumn{2}{@{}l}{\bfseries\small Analysis} \\*[4pt]
Incomplete questionnaires & One blank LifeCLEF 2025 form was excluded. The remaining 90 forms yielded 92 paper-linked quantitative responses. \\[6pt]
Atypical timestamps & No exclusion by timestamp or completion duration; completion times were not collected. \\[6pt]
Statistical correction & Descriptive counts and percentages use full cohort denominators. No nonresponse weighting or inferential tests. A form-level check without multi-paper expansion did not change interpretation. Optional free text informed thematic analysis. \\[6pt]
\end{longtable}
\endgroup

\section{SRQR Checklist}
\label{app:srqr}

\noindent Qualitative reporting follows O\textquotesingle{}Brien et al.~\cite{obrien2014srqr}.

\begingroup
\footnotesize
\renewcommand{\arraystretch}{1.65}
\begin{longtable}{@{}>{\bfseries\raggedright\arraybackslash}p{2.5cm}
                  >{\raggedright\arraybackslash}p{10.5cm}@{}}
\toprule
Checklist item & Study report \\
\midrule
\endhead
\midrule
\multicolumn{2}{r}{\footnotesize\textit{(continued on next page)}}\\
\endfoot
\bottomrule
\endlastfoot
\multicolumn{2}{@{}l}{\bfseries\small Title and abstract} \\*[4pt]
Title & The title identifies author responses to AI-based reviews and the mixed-methods pilot design. \\[6pt]
Abstract & The structured abstract states the purpose, three-cohort sample, descriptive and qualitative methods, principal findings, and interpretive limits. \\[6pt]
\midrule
\multicolumn{2}{@{}l}{\bfseries\small Introduction} \\*[4pt]
Problem formulation & The Background describes variability in peer review, AI-based supplementary feedback, confidentiality concerns, and limited evidence about authors\textquotesingle{} experiences in live workflows. \\[6pt]
Purpose/research questions & The study asks whether authors find AI reviews useful, trust them, perceive complementary issues, use suggestions in revision, and expect notice or consent. \\[6pt]
\midrule
\multicolumn{2}{@{}l}{\bfseries\small Methods} \\*[4pt]
Approach and paradigm & The investigators conducted thematic synthesis informed by reflexive thematic analysis, using semantic coding and primarily inductive theme development within questionnaire topic boundaries. Themes are interpretive accounts rather than fixed categories or frequency estimates. See Methods: Analysis. \\[6pt]
Researcher characteristics and reflexivity & Investigators work in AI and biodiversity research and were involved in organizing workflows, designing prompts, and administering questionnaires. Neither has primary qualitative-methods training. Their investment and organizer roles may shape interpretation and responses. See Methods: Reflexivity. \\[6pt]
Context & LifeCLEF 2025 and LifeCLEF 2026 concern challenge working notes in a recurring CLEF biodiversity initiative; CVWW 2026 is a regional computer-vision workshop. Context, model, prompt, timing, seniority, and procedures co-vary. \\[6pt]
Sampling strategy & Voluntary responses to post-review questionnaires, focused on camera-ready authors. All free text was inspected; 61 substantive extracts from 44 forms informed the themes. No theoretical sampling or saturation criterion was used. \\[6pt]
Ethical issues & Decision V13/2026, 11 May 2026; approval followed LifeCLEF 2025 and CVWW data collection. Voluntary survey submission following an email invitation explaining the research purpose constituted survey consent, separately from manuscript-processing permission. LifeCLEF 2025 used prior notification; CVWW and LifeCLEF 2026 required explicit opt-in before AI processing (Methods and CHERRIES). Names and paper identifiers are excluded from the qualitative analysis table; quotations identify only cohorts. Attribution is separate from analysis. \\[6pt]
Data collection methods & Two optional written questions requested examples of inaccuracies and final reflections. Observed submissions span 8 December 2025--1 February 2026, 13 July--22 August 2026, and 13 January--10 February 2026 for the three cohorts. No interviews or observational data were collected. \\[6pt]
Instruments and technologies & Three Google Forms questionnaires with equivalent substantive domains and minor terminology differences (Appendix~\ref{app:questionnaire}). Inaccuracy examples were elicited using the term ``hallucination,'' which may influence framing. Concerns were also measured categorically. \\[6pt]
Units of study & The quantitative sample comprises 92 responses from 90 completed forms concerning 71 papers. The qualitative corpus contains 23 substantive extracts from 16 LifeCLEF 2025 forms, 20 from 15 LifeCLEF 2026 forms, and 18 from 13 CVWW forms. Each original text is included once, regardless of quantitative weighting. Distinct individuals cannot be enumerated from optional names. \\[6pt]
Data processing & The investigators examined the complete free-text corpus from all three questionnaires. Empty or non-substantive answers were not treated as thematic evidence. Internal form indices link extracts to source text; names and paper identifiers are omitted from the coding table. Quotations were checked against the original wording. \\[6pt]
Data analysis & Corpus familiarization, semantic annotation, candidate-theme development, comparison with all extracts and counterexamples, theme definition, and selection of illustrative quotations. Themes were compared across cohorts and integrated with quantitative findings. The coding record preserves extract-level interpretations; no independent double coding or formal inter-rater statistic was used. \\[6pt]
Trustworthiness & Complete-corpus inspection, traceable extract-level codes, comparison across cohorts, and explicit inclusion of favorable, ambivalent, and divergent views. Quotations were verified against raw text. No member checking or external qualitative audit was conducted. Traceability does not substitute for independent validation. \\[6pt]
\midrule
\multicolumn{2}{@{}l}{\bfseries\small Results} \\*[4pt]
Synthesis and interpretation & Four themes: complementary feedback with selective value; standards, severity, and contextual fit; factual reliability and the limits of evidence checking; author control and accountable use. The analysis distinguishes perceived correctness, usefulness, and authority. \\[6pt]
Links to empirical data & Each theme is supported by cohort-labeled quotations and contextualized accounts. Contrasting views include useful demanding feedback, resistance to venue-provided AI review, support for automatic screening, and consent distinctions between inference and training. Participant claims are not treated as independently verified review assessments. \\[6pt]
\midrule
\multicolumn{2}{@{}l}{\bfseries\small Discussion} \\*[4pt]
Integration, implications, and transferability & The Discussion connects selective usefulness and governance preferences with literature on review assistance, author self-review, and editorial policies. Box 1 proposes conditions for auxiliary use. Cohort contrasts are observational; transferability beyond these computer-science settings is not established. \\[6pt]
Limitations & Self-selection, camera-ready orientation, clustered responses and possible repeated people, self-report, brief optional accounts, investigator/organizer roles, lack of independent review-quality evaluation, and no opportunity for follow-up clarification. See Limitations. \\[6pt]
\midrule
\multicolumn{2}{@{}l}{\bfseries\small Other} \\*[4pt]
Conflicts of interest & No financial competing interests declared. Investigator involvement in the workflows is disclosed in Methods and Declarations. \\[6pt]
Funding & No external funding. \\[6pt]
\midrule
\end{longtable}
\endgroup

\section{Ethics approval}
\label{app:ethics}

\vspace{0.8em}

\noindent\includegraphics[
  page=1,
  width=1\textwidth
]{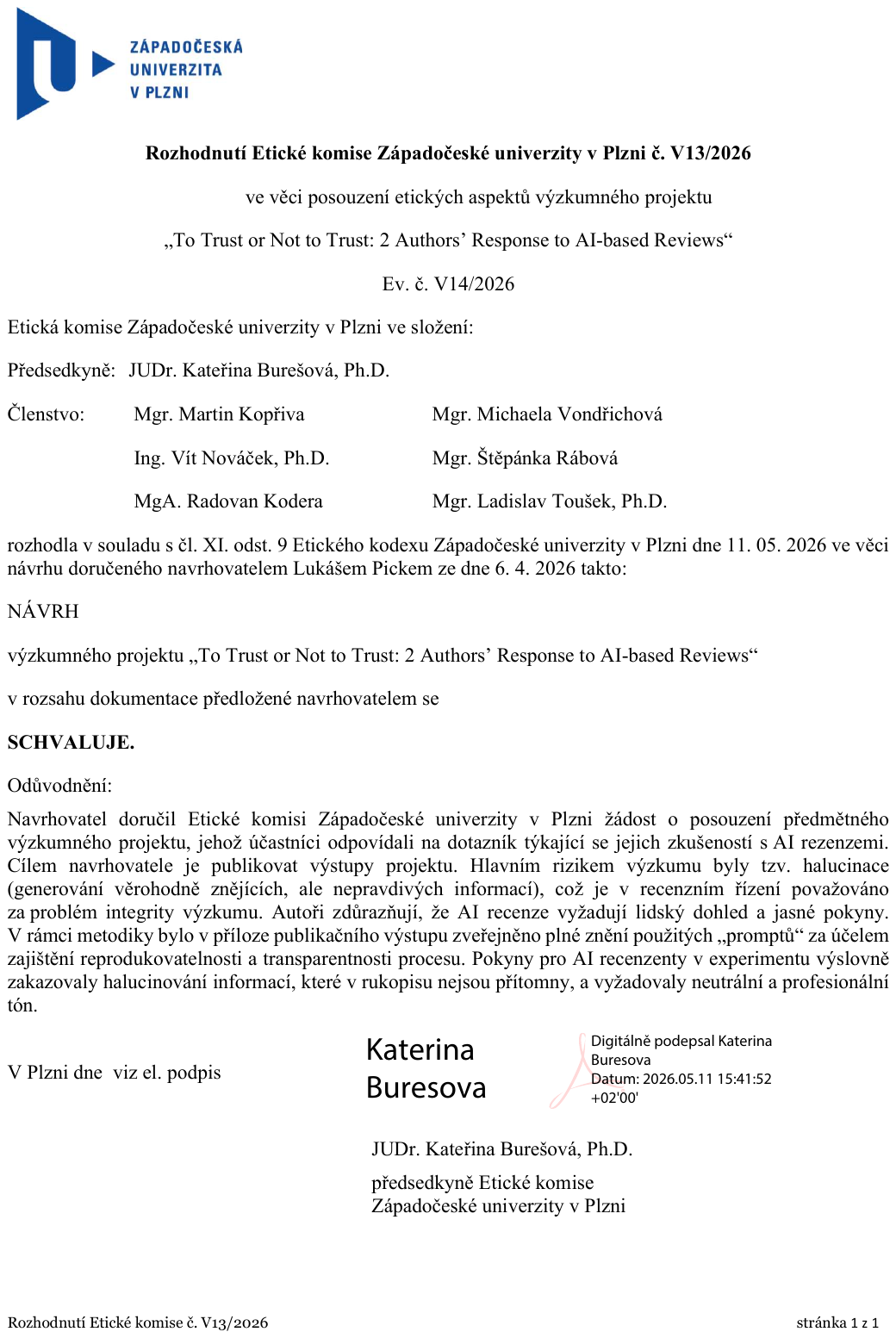}

\end{document}